\documentstyle[12pt]{article}
\def\IR{{\rm I\!R}}
\def\ptl{\partial}
\def\square{\mbox{{$\sqcup$}\llap{$\sqcap$}}}
\def\mapright#1{\mathop{\longrightarrow}\limits_{#1}}
\def\nms{\mathsurround=0pt}
\def\oversim#1#2{\lower 2pt\vbox{\baselineskip 0pt \lineskip 1pt
    \ialign{$\nms#1\hfil##\hfil$\crcr#2\crcr\sim\crcr}}}
\def\ltsim{\mathrel{\mathpalette\oversim<}}
\def\sumprime_#1{\setbox0=\hbox{$\scriptstyle{#1}$}
\setbox2=\hbox{$\displaystyle{\sum}$}
\setbox4=\hbox{${}'\mathsurround=0pt$}
\dimen0=.5\wd0 \advance\dimen0 by-.5\wd2
\ifdim\dimen0>0pt
\ifdim\dimen0>\wd4 \kern\wd4 \else\kern\dimen0\fi\fi
\mathop{{\sum}'}_{\kern-\wd4 #1}}
\newcommand\bmalpha{{\mbox{\boldmath$\alpha$}}}
\newcommand\bmlambda{{\mbox{\boldmath$\lambda$}}}
\renewcommand\theequation{\thesection.\arabic{equation}}
\begin{document}
\title{\bf The Quantized $O(1,2)\big/O(2)\times Z_2$ Sigma Model
Has No Continuum Limit\\
in Four Dimensions.\\
II\@. Lattice Simulation.
\\[-5cm]
{\footnotesize  hep-lat/9205017 \hfill 20 May 1992 \hfill UTREL 92-02}
\\[5cm]
}
\author{\sc Jorge de Lyra,\thanks{Permanent address: \
Universidade de S\~ao Paulo, Instituto de Fisica, Departamento de Fisica
Matem\'atica, C.P. 20516, 01498 S\~ao Paulo S.P., Brazil.}\quad
Bryce DeWitt, See Kit Foong,\thanks{Permanent address: \
Department of Physics, Faculty of Science, Ibaraki University,
Mito 310, Japan.}\\
\sc Timothy Gallivan, Rob Harrington, Arie Kapulkin,\\
\sc Eric Myers, {\rm and} Joseph Polchinski\\[\smallskipamount]
Center for Relativity and Theory Group\\
Department of Physics\\
The University of Texas at Austin\\
Austin, Texas \ 78712}
\date{}
\maketitle\baselineskip18pt
\begin{abstract}\baselineskip12pt
A lattice formulation of the $O(1,2)\big/O(2)\times Z_2$ sigma model is
developed, based on the continuum theory presented in the preceding
paper.  Special attention is given to choosing a lattice action (the
``geodesic'' action) that is appropriate for fields having noncompact
curved configuration spaces.  A consistent continuum limit of the model
exists only if the renormalized scale constant $\beta_R$ vanishes for
some value of the bare scale constant~$\beta$.  The geodesic action has
a special form that allows direct access to the \mbox{small-$\beta$}
limit.  In this limit half of the degrees of freedom can be integrated
out exactly.  The remaining degrees of freedom are those of a compact
model having a \mbox{$\beta$-independent} action which is noteworthy in
being unbounded from below yet yielding integrable averages.  Both the
exact action and the \mbox{$\beta$-independent} action are used to
obtain $\beta_R$ from Monte Carlo computations of field-field averages
(\mbox{2-point} functions) and current-current averages.  Many
consistency cross-checks are performed.  It is found that there is no
value of $\beta$ for which $\beta_R$ vanishes.  This means that as the
lattice cutoff is removed the theory becomes that of a pair of massless
free fields.  Because these fields have neither the geometry nor the
symmetries of the original model we conclude that the
$O(1,2)\big/O(2)\times Z_2$ model has no continuum limit.
\end{abstract}
%%\begin{flushleft}
%%PACS number(s) 11.10Lm, 11.10Gh, 12.25+e, 04.60$-$n\,.
%%\end{flushleft}

\section{Introduction}
In the preceding paper (denoted here by~I) we have outlined the main
theoretical features that would characterize the quantized
$O(1,2)\big/O(2)\times Z_2$ sigma model if it had a continuum limit in
four dimensions.
In this paper we present evidence that the model does not, in fact, have
such a limit.
The equations of paper~I are therefore purely formal and serve only to
motivate our procedures in the lattice simulation.

Because lattice simulations of noncompact sigma models have not been
performed previously, and because there is evidence that the concept of
``universality'' of the scaling critical point may not apply to
noncompact models, we take special care to develop a lattice action that
is appropriate for the $O(1,2)\big/O(2)\times Z_2$ model.
The chief goal of our investigations, which is to determine whether or
not there exists an interacting continuum limit of the model, can be
reached by computing the dimensionless renormalized coupling constant
$\beta_R$ (eq.~(5.12)).
An acceptable continuum limit exits only when $\beta_R$ is zero.
From our numerical simulations we determine $\beta_R$ both from
two-point functions, via the effective action formalism, and from
current algebra.

In addition to performing direct numerical simulations of the
$O(1,2)\big/O(2)\times Z_2$ model we also develop a useful
\mbox{small-$\beta$} or ``strong coupling'' expansion, which permits
half of the degrees of freedom to be integrated out exactly.
The remaining degrees of freedom form a relatively simple
(although somewhat unusual) compact model which is independent of the
bare coupling constant.
Simulations of this model provide yet another way to study the continuum
limit of the full sigma model.

Our numerical results from Monte Carlo simulations of both the full
model and its \mbox{small-$\beta$} limit indicate that $\beta_R$ is
never zero and hence that there is no interacting continuum limit of the
theory.
Our results {\sl are\/} consistent with the existence of a
{\sl noninteracting\/} continuum limit, i.e.,
a continuum limit that is a free field theory.
This result lends support to the popular conjecture that all scalar
field theories in four dimensions are ``trivial''
(in the technical sense of triviality).
However, since this continuum limit has neither the symmetries nor the
interactions of the original classical model we conclude that a
consistent quantum theory of the $O(1,2)\big/O(2)\times Z_2$ model does
not exist.
\section{The massless scalar field.
Lattice action and lattice Fourier transforms.}
To get our bearings on how to introduce the lattice we look first at a
completely soluble system: \ the free massless scalar field.
This system is described by a single real scalar field $\phi$, and its
Euclidean action is
\begin{equation}
S={\textstyle{1\over 2}}\mu^2\int\phi_{,\mu}
\,\phi_{,\mu}\,d^4x\,.\label{1.1}
\end{equation}
(We here drop the subscript $E$ introduced in~I to distinguish the
Euclidean action from the Minkowski one.)
The field $\phi$ is chosen to be dimensionless and the scale factor is
made explicit in order to maintain as close a resemblance as possible to
the sigma models.
The lattice is most simply chosen to be a cubical one having $N^4$ sites
and periodic boundary conditions \mbox{(4-torus)}.
On such a lattice the action (\ref{1.1}) takes the differenced form
\begin{equation}
S={\textstyle{1\over 2}}\mu^2a^2\sum_{\rm links}\left[
\phi\mbox{(one end of link)}-
\phi\mbox{(other end of link)}\right]^2\label{1.2}
\end{equation}
where $a$ is the lattice spacing and the links are between adjacent
sites in the four principal lattice directions, with periodicity
conditions taken into account.

The size, or periodicity distance, of the lattice is
\begin{equation}
L=Na\,,\label{1.3}
\end{equation}
but to the computer this length is irrelevant.
The only adjustable constants in the lattice version of the theory are
the dimensionless quantities $N$ and $\beta$, where $\beta$ is the
coefficient in (\ref{1.2}):
\begin{equation}
\beta\equiv\mu^2a^2\,.\label{1.4}
\end{equation}
If $\mu$ is to be a finite mass scale then $\beta$ must vanish as the
lattice spacing goes to zero.

The system (\ref{1.2}) is best studied by introducing the
{\sl lattice Fourier transform}.
Denote by $\phi_{\alpha\beta\gamma\delta}$ the value of the field $\phi$
at the lattice site having coordinates $\alpha a$, $\beta a$,
$\gamma a$, $\delta a$, with the origin of coordinates taken at a corner
of the periodic cube, the coordinate axes being oriented along the
principal lattice directions, and $\alpha$, $\beta$, $\gamma$, $\delta$
being integers ranging from 0 to $N-1$.
The lattice Fourier transform is defined by
\begin{eqnarray}
\tilde\phi_{klmn}&=&N^{-2}\sum_{\alpha,\beta,\gamma,\delta}
\phi_{\alpha\beta\gamma\delta}\exp
\left[{2\pi i\over N}(k\alpha+l\beta+m\gamma+n\delta)\right]
\,,\label{1.5}\\[\medskipamount]
\noalign{\noindent\rm where $k$, $l$, $m$, $n$ are integers ranging from
${-N}/2+1$ to $N/2$ ($N$~being here assumed to be even).
The original periodic field may be regained by taking the inverse
Fourier transform\medskip}
\phi_{\alpha\beta\gamma\delta}&=&N^{-2}\sum_{k,l,m,n}\tilde\phi_{klmn}
\exp\left[-{2\pi i\over N}(k\alpha+l\beta+m\gamma+n\delta)\right]
\,.\label{1.6}
\end{eqnarray}

It is well known that imposition of periodicity in Euclidean time
corresponds to studying the field $\phi$ in a thermal state at
temperature $T=1/kL$, where $k$ is Boltzmann's constant.
However, since the cutoff energy arising from the lattice discretization
of the system corresponds to a temperature of order $1/ka$, which is $N$
times as big, $T$ may effectively be regarded as vanishing, the thermal
state being indistinguishable from the ground state when $N$ is large.

In order to express the lattice action (\ref{1.2}) in terms of the
Fourier transform $\tilde\phi_{klmn}$, it is convenient to introduce
some abbreviations.
The ordered set of integers $(\alpha,\beta,\gamma,\delta)$ will be
denoted simply by $\bmalpha$ and the ordered set $(k,l,m,n)$ by~$\bf k$.
The following additional abbreviations are obvious:
\begin{eqnarray}
{\bf k}\cdot\bmalpha&=&
k\alpha+l\beta+m\gamma+n\delta\,,\label{1.7}\\[\medskipamount]
\bmalpha+\bmalpha'&=&
(\alpha+\alpha',\beta+\beta',\gamma+\gamma',\delta+\delta')
\,,\label{1.8}\\[\medskipamount]
-\bf k&=&(-k,-l,-m,-n),\mbox{ etc.}\label{1.9}
\end{eqnarray}
Now let the symbol $\bmlambda$ (for links) range over the eight values
$({\pm 1},0,0,0)$, $(0,{\pm 1},0,0)$, $(0,0,{\pm 1},0)$,
$(0,0,0,{\pm 1})$ and regard all integers as defined {\sl modulo\/}~$N$.
Then, taking note of the double counting of links, rewrite expression
(\ref{1.2}) in the form
\begin{eqnarray}
S&=&{\textstyle{1\over 4}}\beta\sum_{\bmalpha,\bmlambda}
\left(\phi_{\smash\bmalpha\vphantom\bmlambda}
     -\phi_{\bmalpha+\bmlambda}\right)^2\nonumber\\[\medskipamount]
&=&{\textstyle{1\over 4}}\beta N^{-4}
\sum_{\bmalpha,\bmlambda,{\bf k},{\bf k}'}
\tilde\phi_{\bf k}\,\tilde\phi_{{\bf k}'}\left[
e^{-(2\pi i/N){\bf k}\cdot\bmalpha}-
e^{-(2\pi i/N){\bf k}\cdot(\bmalpha+\bmlambda)}
\right]\nonumber\\[\medskipamount]
&&\phantom{\beta N^{-4}\sum_{\bmalpha,\bmlambda,{\bf k},{\bf k}'}
\tilde\phi_{\bf k}\tilde\phi_{{\bf k}'}}
\times\left[e^{-(2\pi i/N){\bf k}'\cdot\bmalpha}
           -e^{-(2\pi i/N){\bf k}'\cdot(\bmalpha+\bmlambda)}
\right]\nonumber\\[\medskipamount]
&=&{\textstyle{1\over 4}}\beta\sum_{\bmlambda,\bf k}
\tilde\phi_{\bf k}\,\tilde\phi_{\bf k}{}^*\left[2-
e^{(2\pi i/N){\bf k}\cdot\bmlambda}-
e^{-(2\pi i/N){\bf k}\cdot\bmlambda}\right]\nonumber\\[\medskipamount]
&=&{\textstyle{1\over 2}}\beta\sum_{\bf k}\bar K^2({\bf k})
\left|\tilde\phi_{\bf k}\right|^2\label{1.10}
\end{eqnarray}
where
\begin{eqnarray}
\bar K({\bf k})&\equiv&\left(2\sin{\pi k\over N},
         2\sin{\pi l\over N},2\sin{\pi m\over N},
         2\sin{\pi n\over N}\right)\,,\label{1.11}\\[\medskipamount]
\bar K^2({\bf k})&\equiv&
\bar K({\bf k})\cdot\bar K({\bf k})\,.\label{1.12}
\end{eqnarray}
The last line of (\ref{1.10}) follows from the easily verified identity
\begin{equation}
\sum_\bmlambda\left[1-e^{(2\pi i/N){\bf k}\cdot\bmlambda}\right]=
\sum_\bmlambda\left[1-e^{-(2\pi i/N){\bf k}\cdot\bmlambda}\right]=
\bar K^2({\bf k})\label{1.13}
\end{equation}
and the fact that $\tilde\phi_{-\bf k}=\tilde\phi_{\bf k}{}^*$.

Note that although the physical continuum momentum corresponding to the
integers $k$, $l$, $m$, $n$ is $K({\bf k})\big/a$ where
$K({\bf k})=(2\pi/N)\bf k$, it gets replaced on the lattice by
$\bar K({\bf k})\big/a$, which equals $K({\bf k})\big/a$ only in the
limit $N\to\infty$.
The sine functions appearing in (\ref{1.11}) are lattice artifacts
stemming from the replacement of derivatives by differences.

With the action recast in the form (\ref{1.10}) it is an easy matter to
compute \mbox{2-point} functions, with averages defined as in
eq.~(I.10.5) but adapted for the lattice.
For the free field the integrals in (I.10.5) are Gaussian.
Since the transformation from the variables $\phi_\bmalpha$ to the
variables $\tilde\phi_{\bf k}$ effectively diagonalizes the Gaussian
matrix, and since this transformation is unitary with unit Jacobian, one
may write down the \mbox{2-point} function by inspection:
\begin{equation}
\left<\tilde\phi_{\bf k}\,\tilde\phi_{{\bf k}'}\right>=
{\delta_{{\bf k},-{\bf k}'}\over\beta\bar K^2({\bf k})}\,.\label{1.14}
\end{equation}
The general \mbox{$n$-point} functions are equally easy to obtain.
They vanish when $n$ is odd and are expressible as sums of permuted
products of \mbox{2-point} functions when $n$ is even, which is another
way of saying that all higher correlation functions vanish for Gaussian
probability distributions.

Actual machine computations of the \mbox{2-point} functions are found to
agree very precisely with eq.~(\ref{1.14}) for all values of $\bf k$ and
${\bf k}'$ except $(0,0,0,0)$.
The reason $(0,0,0,0)$ is special
(apart from the fact that expression (\ref{1.14}) becomes singular at
this value) is that $\tilde\phi_{\bf 0}$ represents the
{\sl constant component\/} of $\phi$, being $N^2$ times the average of
$\phi$ over the lattice.
The Euclidean action attains its lower bound, namely zero, whenever
$\phi$ is a constant.
Moreover, the addition of a constant to $\phi$ leaves the action
unchanged, and hence the integrals over the variable
$\tilde\phi_{\bf 0}$ diverge.
On the computer this divergent degree of integration freedom is
eliminated by adding an appropriate constant to $\phi$ at the end of
each Monte Carlo ``sweep'' of the lattice, in order to keep
$\tilde\phi_{\bf 0}$ equal to zero.
This has the consequence that the Monte Carlo average
$\left<\tilde\phi_{\bf 0}{}^n\right>$ vanishes for all positive
integers~$n$.
$\tilde\phi_{\bf 0}$~could, of course, be held fixed at any other value.
Zero is chosen merely for convenience.
This fixing is equivalent to fixing the vacuum state, the vacuum being
here degenerate just as it is in the case of the
$O(1,2)\big/O(2)\times Z_2$ model.

If one takes the lattice Fourier transform of (\ref{1.14})
(remembering $\tilde\phi_{\bf 0}=0$) to obtain the \mbox{2-point}
function directly in ordinary space, one finds that when $N$ is large
and $(\bmalpha-\bmalpha')^2$ is small compared to $N^2$ the
\mbox{2-point} function accurately approximates the continuum Green's
function:
\begin{equation}
\left<\phi_\bmalpha\,\phi_{\bmalpha'}\right>\approx
{1\over 4\pi^2\beta(\bmalpha-\bmalpha')^2},\quad
    N\gg 1,\quad 0<(\bmalpha-\bmalpha')^2\ll N^2\,.\label{1.15}
\end{equation}
Note in particular that this function is spherically symmetric.
For larger values of $(\bmalpha-\bmalpha')^2$, to be sure, the cubical
toroidal lattice structure begins to show itself, but it is nontrivial
that (\ref{1.14}) yields (\ref{1.15}) when $(\bmalpha-\bmalpha')^2$ is
small despite the fact that $\bar K^2({\bf k})$ itself is {\sl not\/}
spherically symmetric.
{\sl A remnant of $O(4)$ symmetry remains in the lattice simulation}.
We shall see later that the momentum-space \mbox{2-point} function of
the $O(1,2)\big/O(2)\times Z_2$ model can be accurately represented by
an expression of the form (\ref{1.14}) multiplied by a nonsingular
slowly varying function of $\bar K^2(k)$.
We regard this as a lattice confirmation of the $O(4)$ symmetry, or
original Lorentz symmetry, of the continuum model.
\setcounter{equation}{0}
\section{A difficulty.
The $O(1,1)\big/Z_2\times Z_2$ model.}
The success achieved in simulating the massless free field on the
computer prompts one to attempt an analogous simulation of the
$O(1,2)\big/O(2)\times Z_2$ model, based on the similarity of expression
(\ref{1.1}) to the part of the action (I.2.5) that does not involve the
Lagrange multiplier.
This similarity suggests that the $O(1,2)\big/O(2)\times Z_2$ lattice
action be chosen in the form
\begin{eqnarray}
S&=&{\textstyle{1\over 2}}\beta\sum_{\rm links}\eta_{ab}
\left[\smash{\phi^a\mbox{(one end of link)}
            -\phi^a\mbox{(other end of link)}}
\vphantom{^b}\right]\nonumber\\[\medskipamount]
&&\phantom{\beta\sum_{\rm links}\eta_a}
\times\left[\phi^b\mbox{(one end of link)}
           -\phi^b\mbox{(other end of link)}\right]\,,\label{2.1}
\end{eqnarray}
with the $\phi^a$ subjected to the constraint
\begin{equation}
\eta_{ab}\,\phi^a\phi^b=-1\label{2.2}
\end{equation}
(cf.~(I.2.8)).
In the case of the $O(3)\big/O(2)$ sigma model, in which $\phi^0$ is
replaced by $\phi^3$, $\eta_{ab}$ by $\delta_{ab}$, and the constraint
by $\phi^a\phi^a=1$, this is exactly the choice of lattice action that
is usually made.
The reason for this is that if the number of dimensions is reduced from
four to three the theory becomes formally identical to the statistical
theory of the Heisenberg ferromagnet, which is interesting to study in
its own right: \
The action becomes proportional to the spin-interaction Hamiltonian,
$\beta$ becomes the coupling constant divided by temperature, and the
functional integral of $e^{-S[\phi]}$ becomes the partition function.

There is, however, a difficulty with the choice (\ref{2.1}).
To show this we digress for a moment by looking at a simpler model, the
$O(1,1)\big/Z_2\times Z_2$ model, which is obtained from the
$O(1,2)\big/O(2)\times Z_2$ model by omitting the field~$\phi^2$.
Equations~(I.2.5) and (\ref{2.2}) continue to hold, but $\eta_{ab}$ is
now given by $\left(\eta_{ab}\right)=\mathop{\rm diag}({-1},1)$.
The configuration space is topologically $\IR$, instead of $\IR^2$, and
can be parameterized by a single variable $s$, where
\begin{equation}
\left.\begin{array}{ll}
\phi^0=\cosh s\\[\medskipamount]
\phi^1=\sinh s.\end{array}\right\}\label{2.3}
\end{equation}
In terms of $s$ the Euclidean action becomes
\begin{equation}
S={\textstyle{1\over 2}}\mu^2\int s_{,\mu}\,s_{,\mu}\,d^4x\label{2.4}
\end{equation}
which is that of the massless free field!

The use of two variables, $\phi^0$ and $\phi^1$, to describe the
massless free field allows us to view it in a new light, and to ask
questions that would not have occurred to us if we had confined our
attention to only one variable, questions that are relevant also for the
$O(1,2)\big/O(2)\times Z_2$ model.
We note first of all that there is a conserved current, which may be
read off directly from eq.~(I.10.20):
\begin{eqnarray}
j_{01\mu}&=&\mu^2\left(\phi_0\,\phi_{1,\mu}
                      -\phi_1\,\phi_{0,\mu}\right)\nonumber\\[\medskipamount]
&=&\mu^2\left[-\cosh s(\sinh s)_{,\mu}
              +\sinh s(\cosh s)_{,\mu}\right]\nonumber\\[\medskipamount]
&=&-\mu^2s_{,\mu}\,.\label{2.5}
\end{eqnarray}
The Fourier transform of this current is simply
\begin{equation}
\tilde\jmath_{01\mu}(p)=i\mu^2p_\mu\,\tilde s(p)\,,\label{2.6}
\end{equation}
which, in view of the well-known free-particle propagator
\begin{eqnarray}
\left<\tilde s(p)\tilde s(p')\right>&=&
{\delta(p+p')\over\mu^2p^2}\,,\label{2.7}\\[\medskipamount]
\noalign{\noindent\rm leads immediately to the following analog of
eq.~(I.10.18):\medskip}
\left<\tilde\jmath_{01\mu}(p)\tilde\jmath_{01\nu}(p')\right>&=&
\mu^2{p_\mu\,p_\nu\over p^2}\delta(p+p')\,.\label{2.8}
\end{eqnarray}
The bare and renormalized Planck masses are seen to be identical for
this model, as is to be expected.

As shown in paper~I the renormalized Planck mass can also in principle
be obtained from any \mbox{2-point} function, for example from
$\left<\phi^1(x)\phi^1(x')\right>$.
This, however, requires the evaluation of $\left<\phi^0\right>$, and
this, in turn, requires introduction of a cutoff such as the lattice
provides.
Let us assume that the correct lattice action for the
$O(1,1)\big/Z_2\times Z_2$ model is expression (\ref{1.2}) with $\phi$
replaced by~$s$.
To obtain $\left<\phi^0\right>$ it is convenient first to
compute $\left<s^2\right>$, using
\begin{equation}
s_\bmalpha=N^{-2}\sumprime_{\bf k}\tilde s_{\bf k}\,
e^{-(2\pi i/N){\bf k}\cdot\bmalpha}\,,\label{2.9}
\end{equation}
where the prime on the summation sign indicates that the term with
${\bf k}=0$ is to be omitted, on the assumption that $\tilde s_{\bf 0}$
is held equal to zero.
Note that holding $\tilde s_{\bf 0}=0$ by adding a constant to the free
scalar field at the end of each Monte Carlo sweep is equivalent to
applying an $O(1,1)$ ``boost'' to the field at the end of each sweep.
This is completely analogous to the procedure that must be adopted in
the case of the $O(1,2)\big/O(2)\times Z_2$ model in order to fix the
vacuum state.
In that case an $O(1,2)$ boost is applied at the end of each sweep in
order to maintain the lattice averages of $\phi^1$ and $\phi^2$ equal to
zero.

Using eq.~(\ref{1.14}), with $\phi$ replaced by $s$, and invoking the
displacement invariance of the theory, one gets
\begin{eqnarray}
\left<s^2\right>&=&\left<s_\bmalpha{}^2\right>=
N^{-4}\sum_\bmalpha\left<s_\bmalpha{}^2\right>=
N^{-4}\sumprime_{{\bf k},{\bf k}',\bmalpha}
\left<\tilde s_{\bf k}\,\tilde s_{{\bf k}'}\right>
e^{-(2\pi i/N)({\bf k}+{\bf k}')\cdot\bmalpha}\nonumber\\[\medskipamount]
&=&N^{-4}\sumprime_{\bf k}
\left<\tilde s_{\bf k}\,\tilde s_{-\bf k}\right>=
F(N)\big/\beta\label{2.10}\\[\medskipamount]
\noalign{\noindent\rm where\medskip}
F(N)&=&N^{-4}\sumprime_{\bf k}
  \bar K^{-2}({\bf k})\,.\label{2.11}\\[\medskipamount]
\noalign{\noindent\rm In the Appendix the following easy estimate of
$F(\infty)$ is derived:\medskip}
F(\infty)&\approx&.15\:.\label{2.12}
\end{eqnarray}
This estimate has been confirmed by explicit summation on the computer.
$F(N)$ is found to converge rapidly to the value .1549\ldots\ for $N$
greater than~4.
This is an example of a stability property that one frequently
encounters in lattice simulations: \
Important quantities often (although not always!) become independent of
$N$ as $N$ is increased.

Because a free field obeys Gaussian statistics one has also
\begin{equation}
\left<s^{2n}\right>={(2n)!\over 2^nn!}\left<s^2\right>^n,\qquad
\left<s^{2n+1}\right>=0\,,\label{2.13}
\end{equation}
whence
\begin{eqnarray}
\left<\phi^0\right>&=&\left<\cosh s\right>
=\sum_{n=0}^\infty{1\over(2n)!}\left<s^{2n}\right>
=\sum_{n=0}^\infty{1\over 2^nn!}\left<s^2\right>^n
=e^{F(N)/2\beta}\label{2.14}\\[\medskipamount]
\noalign{\noindent\rm and\medskip}
\left<\phi^1\right>&=&\left<\sinh s\right>=0\label{2.15}
\end{eqnarray}
(cf.~(I.7.1)).

Suppose now that instead of proceeding as above one chooses for the
lattice action expression (\ref{2.1}) with $a,b\in\{0,1\}$.
In terms of $s$ this action becomes
\begin{equation}
S=2\beta\sum_{\rm links}\sinh^2\textstyle{1\over 2}\left[
s\mbox{(one end of link)}-
s\mbox{(other end of link)}\right]\,.\label{2.16}
\end{equation}
When the link differences are small, expressions (\ref{2.16}) and
(\ref{1.2}) (with $\phi$ replaced by $s$) are approximately equal.
However, in the functional integral the link differences do not stay
small in the continuum limit $a\to 0$, $\beta\to 0$.
Consequently the two actions lead to very different results on the
computer.
This may be seen as follows: \
Let $\Delta s$ be the difference between the values of $s$ at opposite
ends of one of the links in the sum (\ref{2.16}).
The contribution that this link makes to the sum is
$2\beta\sinh^2{1\over 2}\Delta s$.
As long as this contribution is of order 1 or less for every link, the
field will contribute significantly to the functional integrals in
(I.10.5) (with $\phi$ replaced by $s$).
The dominant fields are therefore characterized by the condition
\begin{eqnarray}
\Delta s&\sim&2\sinh^{-1}{1\over\,\sqrt{2\beta}\,}
\,\mapright{\beta\to 0}\,-\ln\beta\,.\label{2.17}\\[\medskipamount]
\noalign{\noindent\rm Since the cross-link differences are of order
$-\ln\beta$ for small $\beta$, one may expect
$\left<\,|s|\,\right>$ to be of similar order:\medskip}
\left<\,|s|\,\right>&\sim&-\ln\beta\label{2.18}\\[\medskipamount]
\noalign{\noindent\rm whence\medskip}
\left<\phi^0\right>&\sim&e^{-\zeta\ln\beta}=\beta^{-\zeta}\label{2.19}
\end{eqnarray}
for some~$\zeta$.
Actual machine computations, using the action (\ref{2.1}), show that
$\zeta$ is approximately equal to .57\,.

The difference between expressions (\ref{2.14})
and (\ref{2.19}) for $\left<\phi^0\right>$ is striking.
According to (\ref{2.19}) $\left<\phi^0\right>$ exhibits a weak
singularity as $\beta\to 0$;
according to (\ref{2.14}) it exhibits an essential singularity!
Another difference shows up in the computation of the
``lattice specific heat,'' which is defined by
\begin{eqnarray}
c_L&=&N^{-4}\left(\left<E^2\right>-\left<\smash E\vphantom{^2}\right>^2\right),
\qquad E\equiv S/\beta\,.\label{3.20}\\[\medskipamount]
\noalign{\noindent\rm Using the action (\ref{1.2})
(with $\phi$ replaced by~$s$) one obtains trivially\medskip}
c_L&=&\mathop{\rm const.}\times\beta^{-2}
\,.\label{3.21}\\[\medskipamount]
\noalign{\noindent\rm On the other hand, computations based on
(\ref{2.1}) are found to yield\medskip}
c_L&=&\mathop{\rm const.}\times\beta^{-1.88(3)}\,.\label{3.22}
\end{eqnarray}
Both (\ref{3.21}) and (\ref{3.22}) show that $\beta=0$ is the
``critical point'' of the theory.
But the critical exponents differ!\,\footnote{It is also worth noting
that $c_L\to\infty$ at the critical point already when $N$ is finite.
With compact models $c_L$ diverges at the critical point only in the
``thermodynamic limit'' $N\to\infty$.}

These results have immediate implications for the
$O(1,2)\big/O(2)\times Z_2$ model.
Obviously one cannot reach trustworthy conclusions about any theory, on
the basis of lattice simulations, unless one can trust the lattice
action.
\setcounter{equation}{0}
\section{The geodesic lattice action}
The sensitivity of the above results to the choice of lattice action
stems from the fact that the fields in noncompact models undergo huge
fluctuations as $\beta\to 0$.
For the $O(1,2)\big/O(2)\times Z_2$ model we should like to use an
action that is as close as possible to the action (\ref{1.2})
(with $\phi$ replaced by~$s$) for the $O(1,1)\big/Z_2\times Z_2$ model.
We first note that the link differences in eq.~(\ref{1.2}) are taken
directly in configuration space (the real line) whereas those in
eq.~(\ref{2.1}) are taken in the embedding space.
In our view one should stay in configuration space.
But unless the configuration space is flat, as it is for the
$O(1,1)\big/Z_2\times Z_2$ model, this criterion is insufficient.
One must also deal with the problem of what ``preferred'' coordinates to
choose, if any.

That this problem may be a critical one can be seen by noting that the
dominant cross-link differences to which the action (\ref{1.2}) gives
rise satisfy
\begin{equation}
\Delta\phi\sim{1\over\,\sqrt\beta\,}={1\over\mu a}\,.\label{3.1}
\end{equation}
Even if one imagines the continuum fields to be $C^\infty$
interpolations of the lattice fields, one sees that in the functional
integrals these fields become wildly discontinuous in the limit
$a\to 0$.
Their derivatives are even more singular, diverging nearly everywhere
like $1/a^2$ as $a\to 0$.

It is easy to see that in $d$ dimensions the derivatives of the fields
diverge nearly everywhere like $a^{-d/2}$ as $a\to 0$.
Thus even when $d=1$, i.e., in the case of ordinary quantum mechanics,
the dominant trajectories are nondifferentiable.
This is a well known feature of {\sl path integration}, but it has
consequences that are sometimes forgotten.
Classically, a system may be described by any convenient variables, and
in a transformation from one set of variables to another, derivatives of
one set are obtained from those of the other by application of the chain
rule.
The chain rule is valid, however, only when the derivatives actually
exist.
In the case of path integration in ordinary quantum mechanics it has
been known for years that the transformation laws for lattice
derivatives do not involve merely simple Jacobians.
Extra terms have to be added to the na\"\i ve finite differences suggested
by the continuum classical action.
In the case of nonlinear sigma models in four dimensions, in which there
are no obviously preferred field variables, we do not have any
already-worked-out theorems to help us in choosing an appropriate
lattice action.
The quantum theory of these models simply does not yet exist as a
coherent discipline.
We are in the position of having to {\sl define\/} the theory by
choosing a specific lattice action {\sl a~priori}.

We have chosen to replace the lattice action (\ref{2.1}) by
\begin{equation}
S={\textstyle{1\over 2}}\beta\sum_{\rm links}\Delta^2\left(
\phi\mbox{(one end of link)},
\phi\mbox{(other end of link)}\right)\label{3.2}
\end{equation}
where $\Delta$ is the geodetic distance (I.3.8) in the configuration
space ({\sl not\/} in the embedding space) between the field values at
the two ends of the link.
This choice has four attractive features:
\begin{enumerate}
\item It can be expressed in terms of any convenient field variables,
being independent of which ones are used.
\item It reduces to (\ref{1.2}) (with $\phi$ replaced by $s$) in the
case of the $O(1,1)\big/Z_2\times Z_2$ model.
\item It is invariant under the actions of the global group, whether
$O(1,1)$ or $O(1,2)$ or whatever.
\item When $d=1$ (ordinary quantum mechanics) it defines, in the
continuum limit, a system whose wave function $\psi$ satisfies the
(almost) obviously correct Schr\"odinger equation
\begin{equation}
i\ptl\psi/\ptl t=\left(-{1\over 2}\square
                       +{1\over 6}R\right)\psi\,,\label{3.3}
\end{equation}
where $\square$ is the covariant Laplacian operator on the configuration
space and $R$ is the curvature scalar.
\end{enumerate}

When $d=1$ the choice (\ref{3.2}) is also obtained if one simply
interpolates classical solutions of the dynamical equations between the
lattice sites and inserts the interpolated $\phi$ into the continuum
action.
For $d>1$ a strict analog of this procedure would require interpolation
of solutions of the classical field equations across lattice
{\sl cells}, but we believe that construction of such an analog would be
unnecessarily complicated and would yield results not significantly
different from those obtained with (\ref{3.2}).
We note also that historical controversies over the term in $R$ in
(\ref{3.3}) are irrelevant here because, for sigma models, $R$ is a
constant and affects only the energy zero point.

The action (\ref{3.2}) can also be used for the $O(3)\big/O(2)$ model.
But in this case the cross-link differences $\Delta\phi^a$ can never be
larger than 2, and simulations we have run show that both (\ref{3.2})
and the conventional action yield results that are qualitatively
similar.
Both lead to phase transitions as $\beta$ is decreased: \
At a certain critical value $\beta_c$, of order unity, the average
$\left<\phi^3\right>$, which plays the role of {\sl magnetization},
drops abruptly to small values (which tend to zero as $N\to\infty$).
The precise value of $\beta_c$ depends on which lattice action is used,
but otherwise there is little difference.

There is a quantity that plays the role of ``magnetization'' for the
$O(1,2)\big/O(2)\times Z_2$ model, namely $\left<\phi^0\right>^{-1}$.
Both (\ref{2.1}) and (\ref{3.2}) cause $\left<\phi^0\right>^{-1}$ to
drop to zero as $\beta\to 0$, but the nature of the dropoff differs
profoundly from one action to the other.
The reason the $O(3)\big/O(2)$ model has a phase transition is that its
configuration space is compact.
$\left<\phi^3\right>$ drops to zero as soon as the functional integral
begins to probe the whole configuration space.
The configuration space of the $O(1,2)\big/O(2)\times Z_2$ model, being
noncompact, is never fully probed as long as $\beta$ is finite.
There is always some residual ``magnetization''
$\left<\phi^0\right>^{-1}$ and there is no phase transition.

These facts have certain implications for the concept of
``universality.''
It is generally believed that, insofar as critical exponents near phase
transitions are concerned, dimensions and symmetries, not details, are
what are important.
Our insistence on choosing the lattice action carefully may seem
superfluous.
In the case of the $O(3)\big/O(2)$ model the choice of lattice action
probably {\sl is\/} irrelevant.
But the $O(1,2)\big/O(2)\times Z_2$ model is different.
It has no phase transition, and important quantities do depend
sensitively on the choice of action.
\setcounter{equation}{0}
\section{The lattice currents}
In the continuum formalism of paper~I the form of the conserved currents
was deduced from the group invariance of the continuum Lagrangian.
The lattice analogs of the currents may be deduced similarly, from the
group invariance of the {\sl lattice Lagrangian}
\begin{equation}
L_\bmalpha=-{1\over 2}\beta\sum_{\bmlambda\mathop{\rm pos}}
\eta_\bmlambda\,\Delta^2\left(\phi_{\smash\bmalpha\vphantom\bmlambda},
                              \phi_{\bmalpha+\bmlambda}\right)\,.\label{4.1}
\end{equation}
Here the summation is over the ``positive'' values of $\bmlambda$,
namely $(1,0,0,0)$, $(0,1,0,0)$, $(0,0,1,0)$ and $(0,0,0,1)$, and the
factor $\eta_\bmlambda$ assumes the values ${-1},1,1,1$ correspondingly.
Introduction of $\eta_\bmlambda$ and the overall minus sign means that
we are working for the moment on a {\sl Minkowski lattice}.
The reason for this is to allow the lattice field equations to have
nontrivial solutions so that there will be something nontrivial to be
conserved!

Reference to eq.~(I.4.1), with the index $\alpha$ replaced by the pair
$ab$, shows that group invariance of $L_\bmalpha$ may be stated in the
form
\begin{eqnarray}
0&=&\delta L_\bmalpha\nonumber\\[\medskipamount]
&=&-{1\over 4}\beta\sum_{\bmlambda\mathop{\rm pos}}\eta_\bmlambda\left[
{\ptl\Delta^2\left(\phi_{\smash\bmalpha\vphantom\bmlambda},
\phi_{\bmalpha+\bmlambda}\right)\over\ptl\phi^i_\bmalpha}
Q^i{}_{ab}\left(\phi_\bmalpha\right)+
{\ptl\Delta^2\left(\phi_{\smash\bmalpha\vphantom\bmlambda},
\phi_{\bmalpha+\bmlambda}\right)\over\ptl\phi^i_{\bmalpha+\bmlambda}}
Q^i{}_{ab}\left(\phi_{\bmalpha+\bmlambda}\right)\right]
\delta\xi^{ab}\,.\nonumber\\
\label{4.2}\\[\medskipamount]
\noalign{\noindent\rm Combining this with the lattice field
equations\medskip}
0&=&{\ptl S\over\ptl\phi^i_\bmalpha}
     ={\ptl\over\ptl\phi^i_\bmalpha}
\sum_{\bmalpha'}L_{\bmalpha'}\nonumber\\[\medskipamount]
&=&-{1\over 2}\beta\sum_{\bmlambda\mathop{\rm pos}}\eta_\bmlambda\left[
{\ptl\Delta^2\left(\phi_{\smash\bmalpha\vphantom\bmlambda},
\phi_{\bmalpha+\bmlambda}\right)\over\ptl\phi^i_\bmalpha}+
{\ptl\Delta^2\left(\phi_{\bmalpha-\bmlambda},
\phi_{\smash\bmalpha\vphantom\bmlambda}\right)\over\ptl\phi^i_\bmalpha}
\right]\label{4.3}\\[\medskipamount]
\noalign{\noindent\rm and remembering that the $\delta\xi^{ab}$ are
arbitrary, one gets\medskip}
0&=&-{1\over 2}\beta\sum_{\bmlambda\mathop{\rm pos}}\eta_\bmlambda\left[
{\ptl\Delta^2\left(\phi_{\smash\bmalpha\vphantom\bmlambda},
\phi_{\bmalpha+\bmlambda}\right)\over\ptl\phi^i_{\bmalpha+\bmlambda}}
Q^i{}_{ab}\left(\phi_{\bmalpha+\bmlambda}\right)-
{\ptl\Delta^2\left(\phi_{\bmalpha-\lambda},
\phi_{\smash\bmalpha\vphantom\lambda}\right)\over\ptl\phi^i_\bmalpha}
Q^i{}_{ab}\left(\phi_\bmalpha\right)\right]\nonumber\\[\medskipamount]
&=&\sum_{\bmlambda\mathop{\rm pos}}\left[
j_{ab}{}^\bmlambda\left(\bmalpha+{1\over 2}\bmlambda\right)-
j_{ab}{}^\bmlambda\left(\bmalpha-{1\over 2}\bmlambda\right)
\right]\label{4.4}
\end{eqnarray}
where
\begin{equation}
j_{ab}{}^\bmlambda\left(\bmalpha+{1\over 2}\bmlambda\right)
\equiv-{1\over 2}\beta\eta_\bmlambda{\ptl\Delta^2
\left(\phi_{\smash\bmalpha\vphantom\bmlambda},
\phi_{\bmalpha+\bmlambda}\right)\over\ptl\phi^i_{\bmalpha+\bmlambda}}
Q^i{}_{ab}\left(\phi_{\bmalpha+\bmlambda}\right)\,.\label{4.5}
\end{equation}
Equation~(\ref{4.4}) is the lattice conservation law and expression
(\ref{4.5}) is the lattice current vector.

Passage to the Euclidean lattice is effected by setting
$\eta_\bmlambda=1$ for all $\bmlambda$ and writing
\begin{eqnarray}
j_{ab\bmlambda}\left(\bmalpha+{1\over 2}\bmlambda\right)&=&-{1\over 2}
\beta{\ptl\Delta^2\left(\phi_{\smash\bmalpha\vphantom\bmlambda},
\phi_{\bmalpha+\bmlambda}\right)\over\ptl\phi^i_{\bmalpha+\bmlambda}}
Q^i{}_{ab}\left(\phi_{\bmalpha+\bmlambda}\right)
\,.\label{4.6}\\[\medskipamount]
\noalign{\noindent\rm In paper~I an alternative and simpler form
(I.4.11 or I.4.17) is given for the currents, based on the action
(I.2.5).
A Lagrange multiplier can also be introduced here, with the result that
expression (\ref{4.6}) gets replaced by its equivalent\medskip}
j_{ab\bmlambda}\left(\bmalpha+{1\over 2}\bmlambda\right)&=&-{1\over 2}
\beta{\ptl\Delta^2\left(\phi_{\smash\bmalpha\vphantom\bmlambda},
\phi_{\bmalpha+\bmlambda}\right)\over\ptl\phi^c_{\bmalpha+\bmlambda}}
G^c{}_{abd}\,\phi^d_{\bmalpha+\bmlambda}\,.\label{4.7}\\[\medskipamount]
\noalign{\noindent\rm Making use of the explicit forms (I.3.7) and
(I.4.15) one finds\medskip}
j_{ab\bmlambda}\left(\bmalpha+{1\over 2}\bmlambda\right)&=&\beta
{\Delta\left(\phi_{\smash\bmalpha\vphantom\bmlambda},
\phi_{\bmalpha+\bmlambda}\right)\over\sinh\Delta\left(
\phi_{\smash\bmalpha\vphantom\bmlambda},
\phi_{\bmalpha+\bmlambda}\right)}\left(
\phi_{\smash{a\bmalpha}\vphantom\bmlambda}\,\phi_{b\bmalpha+\bmlambda}-
\phi_{\smash{b\bmalpha}\vphantom\bmlambda}\,\phi_{a\bmalpha+\bmlambda}
\right)\label{4.8}\\[\medskipamount]
\noalign{\noindent\rm and, in particular,\medskip}
j_{0i\bmlambda}\left(\bmalpha+{1\over 2}\bmlambda\right)&=&-\beta
{\Delta\left(\phi_{\smash\bmalpha\vphantom\bmlambda},
\phi_{\bmalpha+\bmlambda}\right)\over\sinh\Delta\left(
\phi_{\smash\bmalpha\vphantom\bmlambda},
\phi_{\bmalpha+\bmlambda}\right)}\left(
\phi^0_{\smash\bmalpha\vphantom\bmlambda}\phi^i_{\bmalpha+\bmlambda}-
\phi^i_{\smash\bmalpha\vphantom\bmlambda}\phi^0_{\bmalpha+\bmlambda}
\right)\,.\label{4.9}\\[\medskipamount]
\noalign{\noindent\rm The lattice Fourier transforms of these currents,
namely\medskip}
\tilde\jmath_{0i\bmlambda}({\bf k})&=&N^{-2}
\sum_\bmalpha j_{0i\bmlambda}
\left(\bmalpha+{1\over 2}\bmlambda\right)e^{i{\bf k}\cdot
\left(\bmalpha+{1\over 2}\bmlambda\right)}\,,\label{4.10}
\end{eqnarray}
are to be used in the lattice versions of eq.~(I.10.19):
\begin{equation}
\left<\tilde\jmath_{0i\bmlambda}({\bf k})
      \tilde\jmath_{0j\bmlambda'}({\bf k}')\right>
\mapright{{\bf k}\to 0}\beta_R\,\delta_{ij}
{\bmlambda\cdot\bar K({\bf k})
\bmlambda'\cdot\bar K({\bf k})\over\bar K^2({\bf k})}
\delta_{{\bf k},-{\bf k}'}\:,
\qquad\bmlambda\not=\bmlambda'\,.\label{4.11}
\end{equation}
where
\begin{equation}
\beta_R=\mu^2_Ra^2\,.\label{4.13}
\end{equation}
Note: \
Only $\beta_R$, {\sl not\/} $\mu_R$, can be obtained from the computer.
If $\mu_R$ is to remain finite as $a\to 0$ then $\beta_R$
{\sl must vanish in the continuum limit}.
\setcounter{equation}{0}
\section{The small-$\beta$ approximation}
When the lattice action is chosen in the form (\ref{3.2}) the lattice
version of eq.~(I.10.5) may be written
\begin{equation}
\left<A[\phi]\right>={{\displaystyle
\int}A[\phi]\exp\left[-{1\over 4}\beta\sum_{\bmalpha,\bmlambda}\Delta^2
\left(\phi_{\smash\bmalpha\vphantom\bmlambda},
\phi_{\bmalpha+\bmlambda}\right)\right]\displaystyle\prod_\bmalpha
\sinh s_\bmalpha\,ds_\bmalpha\,d\theta_\bmalpha\over{\displaystyle
\int}\exp\left[-{1\over 4}\beta\sum_{\bmalpha,\bmlambda}\Delta^2
\left(\phi_{\smash\bmalpha\vphantom\bmlambda},
\phi_{\bmalpha+\bmlambda}\right)\right]\displaystyle\prod_\bmalpha
\sinh s_\bmalpha\,ds_\bmalpha\,d\theta_\bmalpha}\label{5.1}
\end{equation}
where (cf.~eq.~(I.3.8))
\begin{equation}
\cosh\Delta\left(\phi_{\smash\bmalpha\vphantom\bmlambda},
\phi_{\bmalpha+\bmlambda}\right)
=\cosh s_{\smash\bmalpha\vphantom\bmlambda}\cosh s_{\bmalpha+\bmlambda}
-\sinh s_{\smash\bmalpha\vphantom\bmlambda}\sinh s_{\bmalpha+\bmlambda}
\cos\left(\theta_{\smash\bmalpha\vphantom\bmlambda}-
\theta_{\bmalpha+\bmlambda}\right)\,.\label{5.2}
\end{equation}
When $\beta$ is very small the integrals in (\ref{5.1}) receive
important contributions from a vast range of fields.
For most of these fields $\Delta$ fluctuates over very large values.
If the vacuum is fixed, with the lattice averages of $\phi^1$ and
$\phi^2$ held equal to zero, the $s_\bmalpha$ too fluctuate over large
values, and eq.~(\ref{5.2}) may be accurately replaced by the
approximation
\begin{eqnarray}
\textstyle{1\over 2}e^\Delta&=&\textstyle{1\over 4}\exp
\left(s_{\smash\bmalpha\vphantom\bmlambda}+s_{\bmalpha+\bmlambda}\right)
\left[1-\cos\left(\theta_{\smash\bmalpha\vphantom\bmlambda}-
\theta_{\bmalpha+\bmlambda}\right)\right]\label{5.3}\\[\medskipamount]
\noalign{\noindent\rm whence\medskip}
\Delta&=&s_{\smash\bmalpha\vphantom\bmlambda}
        +s_{\bmalpha+\bmlambda}+\ln\sin^2\textstyle{1\over 2}
\left(\theta_{\smash\bmalpha\vphantom\bmlambda}
     -\theta_{\bmalpha+\bmlambda}\right)\,.\label{5.4}
\end{eqnarray}

The corresponding approximation for the expression standing under the
integral sign of the denominator in eq.~(\ref{5.1}) is
\begin{eqnarray}
&&\exp\left[-{1\over 4}\beta\sum_{\bmalpha,\bmlambda}\Delta^2
\left(\phi_{\smash\bmalpha\vphantom\bmlambda},
\phi_{\bmalpha+\bmlambda}\right)\right]\prod_\alpha\sinh s_\bmalpha
\,ds_\bmalpha\,d\theta_\bmalpha\nonumber\\[\medskipamount]
&&\mapright{\beta\to 0}\,2^{-N^4}\exp\left\{\sum_\bmalpha s_\bmalpha-
{1\over 4}\beta\sum_{\bmalpha,\bmlambda}
\left[s_{\smash\bmalpha\vphantom\bmlambda}+s_{\bmalpha+\bmlambda}+\ln
\sin^2{1\over 2}\left(\theta_{\smash\bmalpha\vphantom\bmlambda}-
\theta_{\bmalpha+\bmlambda}\right)\right]^2\right\}\prod_\bmalpha
ds_\bmalpha\,d\theta_\bmalpha\,.\nonumber\\
\label{5.5}
\end{eqnarray}
The first term in the final exponent represents the effect of the vast
area of configuration space that must be probed by the functional
integrals as $\beta\to 0$.
The linearity (in the $s_\bmalpha$) of this term suggests a shift of
zero point.
Indeed, under the replacement
\begin{equation}
s_\bmalpha={1\over 16\beta}+
{\omega_\bmalpha\over\,\sqrt\beta\,}\label{5.6}
\end{equation}
expression (\ref{5.5}) becomes
\begin{eqnarray}
&&\left({1\over 2}e^{1/32\beta}\right)^{N^4}\exp
\Bigg(-{1\over 4}\sum_{\bmalpha,\bmlambda}
\Bigg\{\left(\omega_{\smash\bmalpha\vphantom\bmlambda}+
\omega_{\bmalpha+\bmlambda}\right)^2+{1\over 4}\ln\sin^2{1\over 2}
\left(\theta_{\smash\bmalpha\vphantom\bmlambda}-
\theta_{\bmalpha+\bmlambda}\right)\nonumber\\[\medskipamount]
&&\phantom{\left({1\over 2}e^{1/32\beta}\right)^{N^4}\exp
\Bigg(-{1\over 4}\sum_{\bmalpha,\bmlambda}\Bigg\{}
+2\sqrt\beta\left(\omega_{\smash\bmalpha\vphantom\bmlambda}+
\omega_{\bmalpha+\bmlambda}\right)\ln\sin^2{1\over 2}
\left(\theta_{\smash\bmalpha\vphantom\bmlambda}-
\theta_{\bmalpha+\bmlambda}\right)\nonumber\\[\medskipamount]
&&\phantom{\left({1\over 2}e^{1/32\beta}\right)^{N^4}\exp
\Bigg(-{1\over 4}\sum_{\bmalpha,\bmlambda}\Bigg\{}
+\beta\left[\ln\sin^2{1\over 2}
\left(\theta_{\smash\bmalpha\vphantom\bmlambda}-
\theta_{\bmalpha+\bmlambda}\right)\right]^2\Bigg\}\Bigg)\prod_\bmalpha
ds_\bmalpha\,d\theta_\bmalpha\,.\label{5.7}
\end{eqnarray}
In the ratio (\ref{5.1}) the factor in front of the above expression is
irrelevant, and if the terms in $\sqrt\beta$ and $\beta$ in the exponent
are discarded, the action takes the effective limiting form
\begin{equation}
S_{\omega,\theta}={1\over 4}\sum_{\bmalpha,\bmlambda}
\left[\left(\omega_{\smash\bmalpha\vphantom\bmlambda}+
\omega_{\bmalpha+\bmlambda}\right)^2+{1\over 4}\ln\sin^2{1\over 2}
\left(\theta_{\smash\bmalpha\vphantom\bmlambda}-
\theta_{\bmalpha+\bmlambda}\right)\right]\,,\label{5.8}
\end{equation}
which is independent of~$\beta$!
Even more remarkable, the variables $\omega_\bmalpha$ are governed by
Gaussian statistics and decouple from the variables~$\theta_\bmalpha$.

The above simple steps appear to give us direct access to the continuum
limit.
We must, of course, check this, through use of the exact lattice action.
The computer results show, in fact, that the approximations embodied in
eqs.~(\ref{5.4}) and (\ref{5.8}) give a good description of the
$O(1,2)\big/O(2)\times Z_2$ model in the \mbox{small-$\beta$} limit.
For the rest of this section and the next section, therefore, we
investigate the implications of the lattice action (\ref{5.8}).

The \mbox{$\omega$-part} of $S_{\omega,\theta}$ is easily expressed in
terms of Fourier components.
Proceeding exactly as in the derivation of eq.~(\ref{1.10}) one finds
\begin{equation}
{1\over 4}\sum_{\bmalpha,\bmlambda}
\left(\omega_{\smash\bmalpha\vphantom\bmlambda}+
\omega_{\bmalpha+\bmlambda}\right)^2={1\over 2}\sum_{\bf k}\hat K^2
({\bf k})\left|\tilde\omega_{\bf k}\right|^2\label{5.9}
\end{equation}
where
\begin{eqnarray}
\hat K({\bf k})&=&\left(2\cos{\pi k\over N},
2\cos{\pi l\over N},2\cos{\pi m\over N},
2\cos{\pi n\over N}\right)\,,\label{5.10}\\[\medskipamount]
\hat K^2({\bf k})&=&16-\bar K^2({\bf k})\,.\label{5.11}
\end{eqnarray}
From this it follows that
\begin{equation}
\left<\tilde\omega_{\bf k}\,\tilde\omega_{{\bf k}'}\right>=
\delta_{{\bf k},-{\bf k}'}\big/\hat K^2({\bf k})\,.\label{5.12}
\end{equation}

Just as in the case of the massless free field there is a zero mode, but
here it corresponds to ${\bf k}={\bf k}_s$ where ${\bf k}_s=
\left({1\over 2}N,{1\over 2}N,{1\over 2}N,{1\over 2}N\right)$
rather than to ${\bf k}=\bf 0$.
On the lattice this mode has the form
\begin{equation}
e^{(2\pi i/N){\bf k}_s\cdot\bmalpha}=(-1)^{|\bmalpha|}\label{5.13}
\end{equation}
where
\begin{equation}
\left|\bmalpha\right|=\left|\alpha\right|+\left|\beta\right|
                     +\left|\gamma\right|+\left|\delta\right|,\qquad
\bmalpha=(\alpha,\beta,\gamma,\delta)\,.\label{5.14}
\end{equation}
Because of its ``spikiness'' expression (\ref{5.13}) will be called the
{\sl spike function}.
Any multiple of the spike function can be added to $\omega_\bmalpha$
without changing the value of the action~$S_{\omega,\theta}$.

The spike mode must be suppressed
(i.e., the sum $\sum_\bmalpha(-1)^{|\bmalpha|}\omega_\bmalpha$ must
always be assumed to vanish) for two reasons:
\begin{enumerate}
\item The spike function does not respect toroidal periodicity when $N$
is odd and hence does not exist as a zero mode in that case.
It is a spurious lattice artifact.
\item If $\sum_\bmalpha(-1)^{|\bmalpha|}\omega_\bmalpha$ were allowed to
drift arbitrarily far from zero then half of the $s_\bmalpha$ of
eq.~(\ref{5.6}) could become small enough to render invalid the
approximation (\ref{5.4}) which led to the introduction of the
$\omega_\bmalpha$ in the first place.
\end{enumerate}

Suppression of the spike mode does not imply that spikiness itself is
removed from the theory.
Indeed it shows up in the \mbox{2-point} function
$\left<\omega_\bmalpha\,\omega_{\bmalpha'}\right>$.
Using double primes on summation signs to indicate that terms with
${\bf k}={\bf k}_s$ are to be omitted, and introducing the abbreviation
\begin{equation}
\hat{\bf k}={\bf k}-{\bf k}_s(\mathop{\rm mod}N),\qquad
\hat K({\bf k})=\bar K(\hat{\bf k})\,,\label{5.15}
\end{equation}
one finds
\begin{eqnarray}
\left<\omega_\bmalpha\,\omega_{\bmalpha'}\right>&=&N^{-4}
\setbox0=\hbox{$\scriptstyle{{\bf k},{\bf k}'}$}
\setbox2=\hbox{$\displaystyle{\sum}$}
\setbox4=\hbox{${}''\mathsurround=0pt$}
\dimen0=.5\wd0 \advance\dimen0 by-.5\wd2
\ifdim\dimen0>0pt
\ifdim\dimen0>\wd4 \kern\wd4 \else\kern\dimen0\fi\fi
\mathop{{\sum}''}_{\kern-\wd4 {\bf k},{\bf k}'}
\left<\tilde\omega_{\bf k}\,\tilde\omega_{{\bf k}'}\right>
e^{-(2\pi i/N)({\bf k}\cdot\bmalpha+{\bf k}'\cdot\bmalpha')}=
\Omega(\bmalpha-\bmalpha')\label{5.16}\\[\medskipamount]
\noalign{\noindent\rm where\medskip}
\Omega(\bmalpha)&=&N^{-4}
\setbox0=\hbox{$\scriptstyle{\bf k}$}
\setbox4=\hbox{${}''\mathsurround=0pt$}
\mathop{{\sum}''}_{\kern-\wd4 \bf k}
{e^{-(2\pi i/N){\bf k}\cdot\bmalpha}\over
\hat K^2({\bf k})}\nonumber\\[\medskipamount]
&=&N^{-4}\sumprime_{\hat{\bf k}}
{e^{-(2\pi i/N)\left(\hat{\bf k}+{\bf k}_s\right)}\over
\bar K^2\left(\hat{\bf k}\right)}=(-1)^{|\bmalpha|}
\Phi(\bmalpha)\,,\label{5.17}
\end{eqnarray}
$\Phi(\bmalpha)$ being the free-field \mbox{2-point} function defined in
the Appendix.
Since $\Phi(\bmalpha)$ is a smooth function $\Omega(\bmalpha)$ is a
spiky function.
Its spikiness indicates that there are correlations between the
$\omega$'s at neighboring sites analogous to spin correlations in
antiferromagnetic matter.

Equation~(\ref{5.17}) is strictly valid only when $N$ is even.
When $N$ is odd it should be replaced by
\begin{equation}
\Omega(\bmalpha)=(-1)^{|\bmalpha|}\Phi_{1/2}(\bmalpha)
\qquad\mbox{($N$ odd)}\label{5.18}
\end{equation}
where $\Phi_{1/2}$ is defined by the sum (\ref{a.2}) in the Appendix but
with the components of $\bf k$ running over half-odd-integral values.
$\Omega(\bmalpha)$ is seen to be spiky for both even and odd~$N$.
On the computer it is convenient always to work with $N$ even.
One can then separate the lattice into {\sl even\/} and {\sl odd\/}
sites, and this permits efficient computer programming through
``vectorization.''

Equations~(\ref{5.6}) and (\ref{5.16}) lead to an easy derivation of
$\left<\phi^0\right>$ in the \mbox{small-$\beta$} limit, patterned on
the corresponding derivation (\ref{2.14}) for the
$O(1,1)\big/Z_2\times Z_2$ model:
\begin{eqnarray}
\left<\phi^0\right>&=&\left<\cosh s\right>=\left<\cosh
\left({1\over 16\beta}+{\omega\over\,\sqrt\beta\,}\right)
\right>\nonumber\\[\medskipamount]
&=&\cosh{1\over 16\beta}\left<\cosh{\omega\over\,\sqrt\beta\,}\right>
  +\sinh{1\over 16\beta}\left<\sinh{\omega\over\,\sqrt\beta\,}\right>
\nonumber\\[\medskipamount]
&=&{1\over 2}e^{1/16\beta}\sum_{n=0}^\infty{1\over(2\beta)^nn!}
\left<\omega^2\right>^n={1\over 2}
e^{\left[{1\over 8}+\Omega(0)\right]/2\beta}\nonumber\\[\medskipamount]
&\mapright{N\to\infty}&\textstyle{1\over 2}
e^{.28/2\beta}\,.\label{5.19}
\end{eqnarray}
The last line follows from
\begin{equation}
\Omega(0)=\Phi(0)=F(N)\mapright{N\to\infty}.1549\ldots\label{5.20}
\end{equation}
(see eqs.~(\ref{5.17}) and (\ref{a.5})).
$\left<\phi^0\right>$ here, like $\left<\phi^0\right>$ for the
$O(1,1)\big/Z_2\times Z_2$ model, has an essential singularity as
$\beta\to 0$, which makes it difficult to evaluate on the computer.
It is, in fact, doubly hard to compute for the following reason: \
In the \mbox{small-$\beta$} limit a histogram of the accumulated values
that $s$ assumes over the lattice has a Gaussian form of width
$\sqrt{\Omega(0)\big/\beta}$ peaked at $1/16\beta=.0625/\beta$.
(See section~8, Figs.~2 and 3, for numerical confirmation of this.)
But the most important contributions to $\left<\phi^0\right>$ come from
values of $s$ near $.14/\beta$, which is way out in the tail of the
Gaussian distribution.
The computed value of $\left<\phi^0\right>$ is extremely sensitive to
the tail of the histogram, and reliable values for it simply cannot be
obtained.
Consequently, eqs.~(I.8.20) to (I.8.23), which are based on the fields
$\phi^a$, do not provide a useful route to the renormalized Planck mass
or, on the computer, to the constant $\beta_R$ of eq.~(\ref{4.13}).
\setcounter{equation}{0}
\section{The fields $\sigma_i$ and
the $\theta$-part of $S_{\omega,\theta}$}
We have found that the Riemann normal variables $\sigma^i$ of
eq.~(I.6.15) {\sl do\/} yield reliable \mbox{2-point} functions on the
computer.
In the \mbox{small-$\beta$} limit the renormalization constant
$Z_\sigma$ of\linebreak
eq.~(I.10.15) becomes
\begin{eqnarray}
Z_\sigma&=&{1\over 2}\left<\left(1+{s\cosh s\over\sinh s}\right)\right>
\mapright{\beta\to 0}\,
{1\over 32\beta}\,,\label{6.1}\\[\medskipamount]
\noalign{\noindent\rm and the renormalized fields (I.10.17) are given
by\medskip}
\sigma^i_R&=&32\beta\sigma^i\,.\label{6.2}
\end{eqnarray}
The renormalized \mbox{2-point} functions therefore have the
\mbox{small-$\beta$} limits
\begin{eqnarray}
&&\left<\sigma^1_{R\bmalpha}\sigma^1_{R\bmalpha'}\right>=(32\beta)^2
\left<s_\bmalpha\,s_{\bmalpha'}\right>\left<\cos\theta_\bmalpha
\cos\theta_{\bmalpha'}\right>\nonumber\\[\medskipamount]
&&\phantom{\left<\sigma^1_{R\bmalpha}\sigma^1_{R\bmalpha'}\right>}
=(32\beta)^2\left<
\left({1\over 16\beta}+{\omega_\bmalpha\over\,\sqrt\beta\,}\right)
\left({1\over 16\beta}+{\omega_{\bmalpha'}\over\,\sqrt\beta\,}\right)
\right>\left<\cos\theta_\bmalpha\cos\theta_{\bmalpha'}\right>
\nonumber\\[\medskipamount]
&&\phantom{\left<\sigma^1_{R\bmalpha}\sigma^1_{R\bmalpha'}\right>}
\mapright{\beta\to 0}\,4\left<\cos\theta_\bmalpha
\cos\theta_{\bmalpha'}\right>\label{6.3}\\[\medskipamount]
\noalign{\noindent\rm and, similarly,\medskip}
&&\left<\sigma^1_{R\bmalpha}\sigma^2_{R\bmalpha'}\right>
\mapright{\beta\to 0}\,4\left<\cos\theta_\bmalpha
\sin\theta_{\bmalpha'}\right>\,,\label{6.4}\\[\medskipamount]
&&\left<\sigma^2_{R\bmalpha}\sigma^2_{R\bmalpha'}\right>
\mapright{\beta\to 0}\,4\left<\sin\theta_\bmalpha
\sin\theta_{\bmalpha'}\right>\,.\label{6.5}\\[\medskipamount]
\noalign{\noindent\rm Owing to the $O(2)$ invariance of the vacuum one
expects expression (\ref{6.4}) to vanish and expressions (\ref{6.3}) and
(\ref{6.5}) to be equal, so that\medskip}
&&\left<\sigma^i_{R\bmalpha}\sigma^j_{R\bmalpha'}\right>
\mapright{\beta\to 0}\delta_{ij}\Theta(\bmalpha-\bmalpha')\label{6.6}
\end{eqnarray}
where
\begin{eqnarray}
\Theta(\bmalpha-\bmalpha')&=&
4\left<\cos\theta_\bmalpha\cos\theta_{\bmalpha'}\right>=
4\left<\sin\theta_\bmalpha\sin\theta_{\bmalpha'}\right>
\nonumber\\[\medskipamount]
&=&2\left<\left(\cos\theta_\bmalpha\cos\theta_{\bmalpha'}
               +\sin\theta_\bmalpha\sin\theta_{\bmalpha'}\right)\right>
=2\left<\cos\left(\theta_\bmalpha-\theta_{\bmalpha'}\right)\right>
\,.\label{6.7}
\end{eqnarray}
The question of the existence of a continuum limit for the
$O(1,2)\big/O(2)\times Z_2$ model is seen to be determined entirely by
the \mbox{$\theta$-part} of the action~$S_{\omega,\theta}$.
The reader should note that the above limiting forms for the
renormalized \mbox{2-point} functions are all independent of~$\beta$.

The \mbox{$\theta$-part} of $S_{\omega,\theta}$ gives rise to the
following unusual average:
\begin{eqnarray}
\left<A[\theta]\right>&=&{\int d\theta A[\theta]
\prod_{\bmalpha,\bmlambda\mathop{\rm pos}}\left|\sin{1\over 2}
\left(\theta_{\smash\bmalpha\vphantom\bmlambda}-
\theta_{\bmalpha+\bmlambda}\right)\right|^{-1/4}\over\int d\theta
\prod_{\bmalpha,\bmlambda\mathop{\rm pos}}\left|\sin{1\over 2}
\left(\theta_{\smash\bmalpha\vphantom\bmlambda}-
\theta_{\bmalpha+\bmlambda}\right)\right|^{-1/4}}
\,,\label{6.8}\\[\medskipamount]
\int d\theta&\equiv&\prod_\bmalpha
\int_0^{2\pi}d\theta_\bmalpha\,.\label{6.9}
\end{eqnarray}
Note that the integrands become singular when the cross-link differences
$\Delta\theta=\theta_{\smash\bmalpha\vphantom\bmlambda}
-\theta_{\bmalpha+\bmlambda}$ vanish.
One therefore expects histograms of the $\Delta\theta$'s to be strongly
peaked around zero.
This is confirmed by the computer results.
(See Fig.~1.)
The singularity of the integrands stems from the fact that the action
$S_{\omega,\theta}$ is unbounded from below (see eq.~(\ref{5.8})).
The question immediately arises why the Monte Carlo procedure
{\sl works\/} when $S_{\omega,\theta}$ is used.
It is shown in the Appendix that the singularities to which
$S_{\omega,\theta}$ gives rise are just soft enough for the functional
integrals to exist.

Another unusual feature of the average (\ref{6.8}) is that its weight,
or ``measure,'' in addition to being invariant under the obvious $O(2)$
transformations
\begin{equation}
\theta_\bmalpha\to\theta_\bmalpha+{\rm constant},\qquad
\theta_\bmalpha\to-\theta_\bmalpha\,,\label{6.10}
\end{equation}
is invariant as well under a set of nonlinear transformations having the
infinitesimal form
\begin{equation}
\delta\theta_\bmalpha=\epsilon_1\cos\theta_\bmalpha
                     +\epsilon_2\sin\theta_\bmalpha\,.\label{6.11}
\end{equation}
The existence of the latter invariance, which is demonstrated in the
Appendix, derives directly from the $O(1,2)$ invariance of the measure
of the average (\ref{5.1}).
The transformation (\ref{6.11}) is easily recognized as an expression of
the well known ``headlight effect'' associated with Lorentz boosts.

When using the ``exact'' action (\ref{3.2}) one must program the
computer to execute $O(1,2)$ boosts at the end of each Monte Carlo sweep
in order to keep the lattice averages of $\phi^1$ and $\phi^2$ equal to
zero and hence to keep the vacuum state from drifting.
When using the \mbox{$\theta$-part} of the action $S_{\omega,\theta}$
one must, for the same reason, program the computer to execute
transformations of the form (\ref{6.11}) after each sweep.
In this case the vacuum is held fixed if the constraints
\begin{equation}
\sum_\bmalpha\cos\theta_\bmalpha=0,\qquad
\sum_\bmalpha\sin\theta_\bmalpha=0\,,\label{6.12}
\end{equation}
which express the $O(2)$ invariance of the vacuum, are maintained.
At the end of each sweep the $\theta_\bmalpha$ will be fairly randomly
distributed from 0 to $2\pi$ (if the lattice is big enough) and will
approximately satisfy the equations
\begin{eqnarray}
N^{-4}\sum_\bmalpha\cos^2\theta_\bmalpha&=&
N^{-4}\sum_\bmalpha\sin^2\theta_\alpha={1\over 2},\qquad
N^{-4}\sum_\bmalpha\sin\theta_\bmalpha
\cos\theta_\bmalpha=0\,.\label{6.13}\\[\medskipamount]
\noalign{\noindent\rm Moreover, eqs.~(\ref{6.12}) will be violated only
slightly.
Defining\medskip}
N^{-4}\sum_\bmalpha\cos\theta_\bmalpha&\equiv&C\ll 1,\qquad
N^{-4}\sum_\bmalpha\sin\theta_\bmalpha\equiv S\ll 1\,,\label{6.14}
\end{eqnarray}
one easily sees that the constraints will, to good accuracy, be
reestablished by the transformation (\ref{6.11}) if $\epsilon_1$ and
$\epsilon_2$ are chosen to be
\begin{equation}
\epsilon_1=-2S,\qquad\epsilon_2=2C\,.\label{6.15}
\end{equation}
This choice has been found to work well on the computer.

We close this section by noting that, in the \mbox{small-$\beta$} limit,
not only the $\sigma$ \mbox{2-point} functions (\ref{6.6}) but also the
current \mbox{2-point} functions (\ref{4.11}) and (5.12) are
determined entirely by the \mbox{$\theta$-part} of the
action~$S_{\omega,\theta}$.
Referring to eqs.~(\ref{4.9}), (\ref{5.4}) and (\ref{5.6}), one easily
obtains the limiting forms
\begin{eqnarray}
j_{01\bmlambda}\left(\bmalpha+{1\over 2}\bmlambda\right)
\mapright{\beta\to 0}-{1\over 16}\,{\cos\theta_{\bmalpha+\bmlambda}
-\cos\theta_{\smash\bmalpha\vphantom\bmlambda}\over
\sin^2{1\over 2}\left(\theta_{\bmalpha+\bmlambda}-
\theta_{\smash\bmalpha\vphantom\bmlambda}\right)}&=&
{1\over 8}\,{\sin{1\over 2}\left(\theta_{\bmalpha+\bmlambda}+
\theta_{\smash\bmalpha\vphantom\bmlambda}\right)\over
\sin{1\over 2}\left(\theta_{\bmalpha+\bmlambda}-
\theta_{\smash\bmalpha\vphantom\bmlambda}\right)}
\,,\label{6.16}\\[\medskipamount]
j_{02\bmlambda}\left(\bmalpha+{1\over 2}\bmlambda\right)
\mapright{\beta\to 0}-{1\over 16}\,{\sin\theta_{\bmalpha+\bmlambda}
-\sin\theta_{\smash\bmalpha\vphantom\bmlambda}\over
\sin^2{1\over 2}\left(\theta_{\bmalpha+\bmlambda}-
\theta_{\smash\bmalpha\vphantom\bmlambda}\right)}&=&
-{1\over 8}\,{\cos{1\over 2}\left(\theta_{\bmalpha+\bmlambda}+
\theta_{\smash\bmalpha\vphantom\bmlambda}\right)\over
\sin{1\over 2}\left(\theta_{\bmalpha+\bmlambda}-
\theta_{\smash\bmalpha\vphantom\bmlambda}\right)}\,,\label{6.17}
\end{eqnarray}
which are independent of the~$\omega_\bmalpha$.
\setcounter{equation}{0}
\section{Computer results}
If one is to use the limiting action $S_{\omega,\theta}$ with
confidence, one must check how well the results obtained with the full
action (\ref{3.2}) tend, when $\beta$ is small, toward those obtained
with~$S_{\omega,\theta}$.
This immediately raises the question: \ How small is {\sl small\/}?

Consider eq.~(\ref{6.3}).
If $\beta$ is not set equal to zero this equation implies
\begin{equation}
\left<\sigma^1_{R\bmalpha}\sigma^1_{R\bmalpha'}\right>=
\left[4+(32)^2\beta\Omega(\bmalpha-\bmalpha')\right]
\left<\cos\theta_\bmalpha\cos\theta_{\bmalpha'}\right>\,,\label{7.1}
\end{equation}
and one sees that the second term in the brackets cannot be ignored
unless
\begin{equation}
\beta\ll 4(32)^{-2}\!\big/\Omega(0)=.025~.\label{7.2}
\end{equation}
Such small values for $\beta$ have the potential of severely taxing the
ability of the computer.
In early runs attempts were made to compute
$\left<\phi^i_{R\bmalpha}\phi^j_{R\bmalpha'}\right>$ at
$\beta\ltsim.002$, without success.
The cause of the failure was the same as that which makes the
computation of $\left<\phi^0\right>$ impractical, as explained at the
end of section~6.
On the computer the difficulty can masquerade as
{\sl critical slowing down}, causing large autocorrelation intervals,
drifting averages, etc.
But in fact critical slowing down has nothing to do with it.
Computations of $\left<\sigma^i_{R\bmalpha}\sigma^j_{R\bmalpha'}\right>$
have turned out to generate autocorrelation intervals of less than
20~sweeps, even for $\beta$'s as small as $10^{-4}$.
Moreover, the results have been found to be both reliable and accurate.

A simple measure of the validity of the \mbox{small-$\beta$}
approximation is the quantity
\begin{equation}
\left<{\,|\Delta'-\Delta|\,\over\Delta}\right>\,,\label{7.3}
\end{equation}
where $\Delta$ is the exact cross-link geodetic distance, $\Delta'$ is
the approximation to it given by eq.~(\ref{5.4}), and the average is
over links and sweeps.
At $\beta=10^{-3}$ expression (\ref{7.3}) is found to be already less
than $10^{-7}$, so the \mbox{small-$\beta$} approximation is indeed a
good one.

More detailed measures of the quality of the \mbox{small-$\beta$}
approximation are provided by histograms of the $s_\bmalpha$ and of the
cross-link differences $\Delta\theta=
\theta_{\smash\bmalpha\vphantom\bmlambda}-\theta_{\bmalpha+\bmlambda}$.
Figure~1 shows the $\Delta\theta$ histograms for both the exact lattice
action (\ref{3.2}), at $\beta=10^{-4}$ and $N=10$, and the
\mbox{$\theta$-part} of the approximate action $S_{\omega,\theta}$, at
$N=10$.
The two curves, which were compiled by accumulating all cross-link
differences for 60,000 sweeps, are indistinguishable.
Figures~2 and 3 show the $s_\bmalpha$ histograms for the exact action,
at $\beta=10^{-3}$ and $10^{-4}$ with $N=10$, superimposed upon the
theoretical Gaussian curve derived from eq.~(\ref{5.6}) and the
\mbox{$\omega$-part} of the action~$S_{\omega,\theta}$.
Again the agreement is excellent.

It should be stated that during the runs at $\beta\le 10^{-3}$ the spike
mode (see section~6) began to appear spontaneously and randomly, in the
fields generated by the Monte Carlo procedure.
At these values of $\beta$ this spurious mode was forcibly removed from
the field at the end of each sweep after the vacuum-fixing boost was
applied.
As remarked in section~6, suppression of the spike mode does not remove
spikiness itself from the theory.
(See the discussion of the $\sigma$ \mbox{2-point} function below.)

It should also be stated that during the \mbox{small-$\beta$} runs the
vacuum-fixing boosts applied at the end of each sweep were chosen so as
to enforce the conditions (\ref{6.12}) rather than the conditions
\begin{equation}
\sum_\bmalpha\phi^1_\bmalpha=0,\qquad
\sum_\bmalpha\phi^2_\bmalpha=0\,.\label{7.4}
\end{equation}
Because of the huge values assumed by the $\phi^i_\bmalpha$
(${\sim 10^{27\pm 10}}$ when $\beta=10^{-3}$ and
${\sim 10^{273\pm 33}}$ when $\beta=10^{-4}$) enforcement of (\ref{7.4})
is impractical at these~$\beta$'s.
We {\sl could\/} have imposed the alternative conditions
\begin{equation}
\sum_\bmalpha\sigma^1_\bmalpha=0,\qquad
\sum_\bmalpha\sigma^2_\bmalpha=0\,,\label{7.5}
\end{equation}
but we felt that, to make comparisons, it was best to uniformly impose
conditions (\ref{6.12}), which are the only ones available when the
action $S_{\omega,\theta}$ is used.
It turns out that enforcement of (\ref{6.12}) causes conditions
(\ref{7.5}) also to be satisfied to better than .03\% accuracy provided
$N$ is big enough $({\ge 10})$ so that a ``snapshot'' of the lattice
after each boost provides a fair sampling of field values.

The results displayed in Figs.~1, 2 and 3 give us confidence
in proceeding to the computation of~$\beta_R$.
We discuss first the determination of $\beta_R$ based on the lattice
version of eq.~(I.10.16):
\begin{equation}
\left<\tilde\sigma^i_{R\bf k}\tilde\sigma^j_{R{\bf k}'}\right>
={\delta_{ij}\,\delta_{{\bf k},-{\bf k}'}\over
\beta_R\,\bar K^2({\bf k})+\cdots}\:.\label{7.6}
\end{equation}
Here $\tilde\sigma^i_{R\bf k}$ is the lattice Fourier transform of the
field $\sigma^i_{R\bmalpha}=Z^{-1}_\sigma\sigma^i_\bmalpha$.
Note that the determination of $\beta_R$ requires a computation of the
renormalization constant $Z_\sigma$ (eqs.~(I.10.15) and (\ref{6.1})) as
well as the $\sigma$ \mbox{2-point} function.
According to eq.~(\ref{7.6}) $\beta_R$ is given by the intercept
(at ${\bf k}=0$) of the function $2/\bar K^2({\bf k})
\left<\tilde\sigma^i_{R\bf k}\tilde\sigma^i_{R\bf k}{}^*\right>$.
Figure~4 shows plots of this function for $N=10$ and various values of
$\beta$, as well as of the corresponding function for the action
$S_{\omega,\theta}$ (eqs.~(\ref{6.6}) and (\ref{8.7})).
At the smaller values of $\beta$ and at low momenta the curves fall
right on top of one another.
The differences between the curves at higher momenta is explained by the
spikiness of the function
$\left<\sigma^i_{R\bmalpha}\sigma^i_{R\bmalpha'}\right>$ arising from
$\Omega(\bmalpha-\bmalpha')$ in eq.~(\ref{7.1}).
The spikiness is a short-wavelength phenomenon that is picked up by the
Fourier transforms.
It goes away as $\beta\to 0$.

Figure~5 shows a plot of $\beta_R$ vs.~$\beta$ as determined from the
intercepts of the curves shown in Fig.~4.
It is on this plot that we base our conclusion that the quantized
$O(1,2)\big/O(2)\times Z_2$ sigma model has no continuum limit in four
dimensions.
The key fact is that $\beta_R$ {\sl vanishes nowhere, not even at\/}
$\beta=0$.
In view of the definition of $\beta_R$ (eq.~(5.12)) this implies
that the renormalized Planck mass $\mu_R$ must become infinite in the
continuum limit $a\to 0$.

Figure~6 shows a plot of $\beta_R$ vs.~$N$ for the action
$S_{\omega,\theta}$, i.e., for $\beta=0$.
It is evident that $\beta_R$ tends to a constant value as $N$ increases
past 10.
Its value on an {\sl infinite\/} lattice is therefore effectively known,
and one can pass to the continuum limit $a\to 0$ without worrying that
the size of the ``universe'' (the lattice \mbox{4-torus}) is
simultaneously collapsing to zero.
The nonvanishing of $\beta_R$, and the blowup of $\mu_R$, is {\sl not\/}
an artifact of a vanishing universe.

If a similar result (infinite $\mu_R$) is eventually found in
conventional quantum gravity we shall be forced to conclude that the
theory is not viable, because the classical theory of Einstein cannot
emerge from it as a low-energy limit.
In the case of the $O(1,2)\big/O(2)\times Z_2$ model the blowup of
$\mu_R$ implies that there can be no low-energy ``classical'' regime in
which the physics is described by an action functional having the form
of the first term on the right side of eq.~(I.8.18).
It is in this sense that we say that the quantized
$O(1,2)\big/O(2)\times Z_2$ model has no continuum limit.

It may occasionally be useful to adopt an alternative viewpoint by
saying that the continuum limit exists but is trivial.
However, this viewpoint can be misleading on two accounts:
\begin{enumerate}
\item It reinforces the current folklore which claims that all scalar
field theories in four dimensions are trivial.
The claim is suspect because it {\sl is\/} just folklore.
No valid argument exists that would have given us knowledge
{\sl a~priori\/} that $\mu_R$ must blow up in the continuum limit.
Each scalar theory must be investigated on its own.
For example, despite the amount of research that has been done on the
$O(3)\big/O(2)$ model, it is still unknown whether a continuum limit
exists for it when the action (\ref{3.2}) is used.
If $\beta_R$ does vanish anywhere for this model one presumes that it
will do so at the phase-transition point $\beta=\beta_c$, but no one
knows.
\item While it is true that the $O(1,2)\big/O(2)\times Z_2$ model
becomes trivial in the continuum limit, in the sense that the
renormalized ``coupling constant'' $\mu^{-2}_R$ vanishes, it is also
true that in this limit the model is no longer
$O(1,2)\big/O(2)\times Z_2$.
The configuration space becomes flat and the invariance group consists
of the isometries of the plane, not the elements of $O(1,2)$.
Therefore, although the continuum limit ``exists,'' the model does not.
\end{enumerate}

We remark that the continuum limit may properly be regarded as located
at $\beta=0$, not only because the na\"\i ve continuum limit is located
there (with the bare Planck mass $\mu$ held at a fixed finite value) but
also because the lattice specific heat diverges there, making $\beta=0$
a critical point of the theory.
This may be seen by introducing the ``partition function''
\begin{equation}
Z(\beta)=\int e^{-\beta E[\phi]}\mu[\phi]\,d\phi\label{8.7}
\end{equation}
and noting that eq.~(\ref{3.20}) is equivalent to
\begin{equation}
c_L=N^{-4}\left[{1\over Z}\,{d^2Z\over d\beta^2}
           -{1\over Z^2}\left({dZ\over d\beta}\right)^2\right]
\,.\label{8.8}
\end{equation}
In the \mbox{small-$\beta$} limit the partition function is dominated by
the first factor in expression (\ref{5.7}), whence
\begin{eqnarray}
Z&=&\mathop{\rm const.}\times e^{N^4\!/32\beta}
\,.\label{8.9}\\[\medskipamount]
\noalign{\noindent\rm This immediately yields\medskip}
c_L&=&\mathop{\rm const.}\times\beta^{-3}\,,\label{8.10}
\end{eqnarray}
which diverges at $\beta=0$.
The critical exponent, ${-3}$, in eq.~(\ref{8.10}) has been confirmed by
direct computation of (\ref{3.20}) for the exact action (\ref{3.2}).
Note that this critical exponent differs from that of the massless free
field (eq.~(\ref{3.21})).

The ``triviality'' of the continuum limit of the
$O(1,2)\big/O(2)\times Z_2$ model, which is entirely due to the
nonvanishing of $\beta_R$, implies nothing about the shape of the
$\beta=0$ curve in Fig.~4.
Since this curve is plotted effectively as a function of
$p^2\!\big/\mu_R{}^2$, and since $\mu_R{}^2$ diverges, finite physical
values of $p^2$ are all located at the left-hand intercept.
It is therefore quite remarkable that this curve is nevertheless very
nearly a straight horizontal line.
Such a result was by no means obvious {\sl a~priori}.
Suppose one were to present a mathematician with eq.~(\ref{6.8}) without
telling him where it came from.
How could he possibly know in advance that the average (\ref{6.7}),
defined by (\ref{6.8}), would be very nearly the \mbox{2-point} function
for a pair of free fields?

There is a consistency check on this curious result.
Once one knows that the average (\ref{6.7}) is very nearly proportional
to the \mbox{2-point} function $\Phi(\bmalpha-\bmalpha')$ for the free
field, one may use the proportionality constant to define $\beta_R$ at
$\beta=0$:
\begin{equation}
\Theta(\bmalpha-\bmalpha')\approx\beta^{-1}_R
  \Phi(\bmalpha-\bmalpha')\,.\label{7.11}
\end{equation}
But from eq.~(\ref{6.7}) it is obvious that $\Theta(0)=2$.
Moreover
\begin{eqnarray}
\Phi(0)=F(N)&\mapright{N\to\infty}&
.1549\,,\label{7.12}\\[\medskipamount]
\noalign{\noindent\rm whence it must follow that\medskip}
\beta_R\approx{1\over 2}F(N)&\mapright{N\to\infty}&
.0775\,,\label{7.13}
\end{eqnarray}
which agrees very well with Fig.~6.

It is useful to sketch the procedure that we should have had to adopt if
things had turned out differently and $\beta_R$ were to vanish
for some value of $\beta$.
To begin with we should have tried to match the $\sigma$ \mbox{2-point}
function of the continuum theory to an Ansatz of the form
\begin{equation}
\left<\tilde\sigma^i_R(p)\tilde\sigma^j_R(p')\right>=
{\delta_{ij}\,\delta(p+p')\over\mu^2_Rp^2+\left(\alpha/\mu^2_R\right)
p^6\ln\left(p^2\!/\lambda\mu^2_R\right)}\label{7.7}
\end{equation}
or some variant thereof.
Here $\alpha$ and $\lambda$ are dimensionless fitting parameters
independent of~$\mu_R$.
Equation~(\ref{7.7}) would be a universal formula, with $\mu_R$ merely
setting the scale.

The lattice analog of (\ref{7.7}) is
\begin{equation}
\left<\tilde\sigma^i_{R\bf k}\tilde\sigma^j_{R{\bf k}'}\right>=
{\delta_{ij}\,\delta_{{\bf k},-{\bf k}'}\over\beta_R
\,\bar K^2+\left(\alpha/\beta_R\right)\bar K^6\ln
\left(\bar K^2\!/\lambda\beta_R\right)}\:.\label{7.8}
\end{equation}
This formula becomes meaningless when $\beta_R$ vanishes, and one might
therefore be tempted to conclude that, since the average (\ref{6.7}) is
independent of $\beta$, it was obvious, as soon as eqs.~(\ref{6.6}) and
(\ref{6.7}) were written down, that $\beta_R$ cannot vanish.
But this would be too hasty a conclusion.
Every finite lattice has associated with it an ``infrared'' mass
\begin{equation}
\mu'=1/L=1/Na\,.\label{7.9}
\end{equation}
The corresponding lattice quantity is
\begin{equation}
\beta'=\mu'{}^2a^2=N^{-2}\,,\label{7.10}
\end{equation}
which, because it remains finite as $a$ goes to zero, could quite
possibly mask $\beta_R$ in the limit.
If the real $\beta_R$ had turned out to vanish then one would have found
that the {\sl effective\/} $\beta_R$ defined by eq.~(\ref{7.6}) was
proportional to~$N^{-2}$.
One might have regarded any hope for such a result as foredoomed, in
view of the insensitivity of most averages to $N$, once $N$ is bigger
than 6 or~8.
But eq.~(\ref{6.8}) defines an unusual kind of average, and we had no
advance knowledge of its properties.

We turn finally to the determination of $\beta_R$ from the
current-current average (\ref{4.11}).
Because of large scatter in the individual values of
$\tilde\jmath_{0i\bmlambda}({\bf k})
\tilde\jmath_{0i\bmlambda'}(k)^*$ we found that it was not possible to
use eq.~(\ref{4.11}) at values of $\beta$ below $10^{-2}$.
But down {\sl to\/} that value the agreement between the values of
$\beta_R$ obtained from the $\sigma$ \mbox{2-point} function and those
obtained from (\ref{4.11}) is good.
Indeed, Fig.~7 shows that the agreement is remarkably good.
Because the determination of $\beta_R$ from (\ref{4.11}) avoids
computation of the renormalization constant $Z_\sigma$, this good
agreement is far from trivial.
It is a convincing demonstration of the consistency of the formal
arguments presented in paper~I\@.
It makes one believe that there is something to quantum field theory
after all!

In closing we note yet another consistency check on the formalism.
Equations~(I.9.2), (I.9.3), (I.9.4) and (I.9.7) together imply
\begin{equation}
\left<\mathop{\rm vac}\right|T\left(\sigma^i(x)j_{0j}{}^\mu(x')\right)
\left|\mathop{\rm vac}\right>={\delta_{ij}\,Z_\sigma\over(2\pi)^4}\int
{p^\mu\over p^2}e^{ip\cdot(x-x')}\,d^4p\,.\label{7.14}
\end{equation}
Replacing $\sigma^i$ by $\sigma^i_R$, passing to Euclidean space and
taking Fourier transforms, we get
\begin{equation}
\left<\tilde\sigma^i_R(p)\tilde\jmath_{0j}{}^\mu(p)\right>
=i\delta_{ij}{p^\mu\over p^2}\delta(p+p')\,.\label{7.15}
\end{equation}
The lattice version of this equation yields
\begin{equation}
\left|\left<\tilde\sigma^i_{R\bf k}
\tilde\jmath_{0i\bmlambda}({\bf k})^*\right>\right|=2
{\bmlambda\cdot\bar K({\bf k})\over\bar K^2({\bf k})}\,,\label{7.16}
\end{equation}
which is a universal function, the same for all values of~$\beta$.
Equation (\ref{7.16}) has been checked at $N=4$ for all values of
$\bf k$ and $\bmlambda$ and for many values of~$\beta$.
We have found it to be satisfied to 5\% accuracy for values
of $\beta$ down to .05, and to 20\% accuracy at $\beta=.02\,$.
At lower values of $\beta$ large scatter makes checking impossible.
Although eq.~(\ref{7.16}) gives no information about the continuum
limit, it is nevertheless a remarkable equation.
Its left-hand side depends on $Z_\sigma$ and $\beta$;
its right-hand side depends on neither.
Its ``experimental'' verification gives us one more reason to believe
that we have done the right things.

\vspace{\bigskipamount}
This work has been supported in part by Grants PHY8919177, PHY9009850,
and\linebreak
PHY9120042 from the National Science Foundation, and by
the Robert A. Welch Foundation.
Computer services have been provided by
the Pittsburgh Supercomputing Center on Grant No.~PHY880071P,
the National Center for Supercomputing Applications, Champaign-Urbana,
and The University of Texas System
Center for High Performance Computing, Austin.
\clearpage
\renewcommand\theequation{A.\arabic{equation}} \setcounter{equation}{0}
\section*{Appendix}
\subsection*{Estimating $F(N)$}
To estimate $F(N)$ use the approximation (\ref{1.15}).
The exact equation for the \mbox{2-point} function is
\begin{equation}
\left<\phi_\bmalpha\,\phi_{\bmalpha'}\right>=
{1\over\beta}\Phi(\bmalpha-\bmalpha')\label{a.1}
\end{equation}
where (cf.~eq.~(\ref{1.14}))
\begin{eqnarray}
\Phi(\bmalpha)&=&N^{-4}\sumprime_{\bf k}
{e^{-(2\pi i/N){\bf k}\cdot\bmalpha}\over\bar K^2({\bf k})}
\,.\label{a.2}\\[\medskipamount]
\noalign{\noindent\rm Evidently\medskip}
\Phi({\bf 0})&=&F(N)\,,\label{a.3}\\[\medskipamount]
\noalign{\noindent\rm and, from eq.~(\ref{1.15}),\medskip}
\Phi(\bmlambda)&\approx&{1\over 4\pi^2\bmlambda^2}={1\over 4\pi^2}
=.025\ldots\mbox{for all $\bmlambda$}\,.\label{a.4}
\end{eqnarray}
Now use the fact that $\Phi$ satisfies a simple equation involving the
lattice Laplacian:
\begin{eqnarray}
&&{1\over 2}\sum_\bmlambda\left[-\Phi(\bmalpha+\bmlambda)
                +2\Phi(\bmalpha)-\Phi(\bmalpha-\bmlambda)\right]
\nonumber\\[\medskipamount]
&&{}={1\over 2}N^{-4}\sum_\bmlambda\,\sumprime_{\bf k}\left[-
e^{-(2\pi i/N){\bf k}\cdot(\bmalpha+\bmlambda)}+2
e^{-(2\pi i/N){\bf k}\cdot\bmalpha}-
e^{-(2\pi i/N){\bf k}\cdot(\bmalpha-\bmlambda)}\right]
\bigg/\bar K^2({\bf k})\nonumber\\[\medskipamount]
&&{}=N^{-4}\sumprime_{\bf k}e^{-(2\pi i/N){\bf k}\cdot\bmalpha}
=\delta_{\bmalpha\bf 0}-N^{-4}\label{a.5}
\end{eqnarray}
in which eq.~(\ref{1.13}) is used in passing to the last line.
Setting $\bmalpha=\bf 0$ in (\ref{a.5}) and remembering that $\bmlambda$
ranges over eight different values, get
\begin{eqnarray}
1-N^{-4}&=&8\left[\Phi({\bf 0})-\Phi(\bmlambda)\right]
\approx 8\left[F(N)-.025\right]\,,\nonumber\\[\medskipamount]
\noalign{\noindent\rm whence\medskip}
F(N)&\approx&.025+{1\over 8}\left(1-N^{-4}\right)
\mapright{N\to\infty}\,.15~.\label{a.6}
\end{eqnarray}
\subsection*{Invariance of the measure of the average (\ref{6.8})}
The measure may be written in the form
\begin{equation}
\prod_\bmalpha d\theta_\bmalpha\prod_\bmlambda\left[\sin^2{1\over 2}
\left(\theta_{\smash\bmalpha\vphantom\bmlambda}-
\theta_{\bmalpha+\bmlambda}\right)\right]^{-1/16}\,.\label{a.7}
\end{equation}
Setting $\theta'_\bmalpha=\theta_\bmalpha+\delta\theta_\bmalpha$ one
finds for the change in $d\theta_\bmalpha$ under the infinitesimal
transformation (\ref{6.11})
\begin{eqnarray}
\delta\,d\theta_\bmalpha&=&d\theta'_\bmalpha-d\theta_\bmalpha=\left[
\left(\ptl\theta'_\bmalpha\big/\ptl\theta_\bmalpha\right)-1\right]
\,d\theta_\bmalpha=
\left(\ptl\delta\theta_\bmalpha\big/\ptl\theta_\bmalpha\right)
\,d\theta_\bmalpha\nonumber\\[\medskipamount]
&=&\left(-\epsilon_1\sin\theta_\bmalpha
         +\epsilon_2\cos\theta_\bmalpha\right)
\,d\theta_\alpha\label{a.8}\\[\medskipamount]
\noalign{\noindent\rm and hence\medskip}
\delta\prod_\bmalpha d\theta_\bmalpha&=&
\left(\prod_\bmalpha d\theta_\bmalpha\right)\sum_\bmalpha\left(
-\epsilon_1\sin\theta_\bmalpha
+\epsilon_2\cos\theta_\bmalpha\right)\,.\label{a.9}
\end{eqnarray}
Moreover,
\begin{eqnarray}
&&\delta\left[\sin^2{1\over 2}\left(
\theta_{\smash\bmalpha\vphantom\bmlambda}-
\theta_{\bmalpha+\bmlambda}\right)\right]^{-1/16}
\nonumber\\[\medskipamount]
&&{}=-{1\over 16}\left[\sin^2{1\over 2}\left(
\theta_{\smash\bmalpha\vphantom\bmlambda}+
\theta_{\bmalpha+\bmlambda}\right)\right]^{-1/16}{\cos{1\over 2}\left(
\theta_{\smash\bmalpha\vphantom\bmlambda}-
\theta_{\bmalpha+\bmlambda}\right)\over\sin{1\over 2}\left(
\theta_{\smash\bmalpha\vphantom\bmlambda}-
\theta_{\bmalpha+\bmlambda}\right)}\left(\delta
\theta_{\smash\bmalpha\vphantom\bmlambda}-\delta
\theta_{\bmalpha+\bmlambda}\right)\nonumber\\[\medskipamount]
&&{}=-{1\over 16}\left[\sin^2{1\over 2}\left(
\theta_{\smash\bmalpha\vphantom\bmlambda}+
\theta_{\bmalpha+\bmlambda}\right)\right]^{-1/16}\bigg[-2\epsilon_1
\cos{1\over 2}\left(\theta_{\smash\bmalpha\vphantom\bmlambda}-
\theta_{\bmalpha+\bmlambda}\right)\sin{1\over 2}\left(
\theta_{\smash\bmalpha\vphantom\bmlambda}+
\theta_{\bmalpha+\bmlambda}\right)\nonumber\\[\medskipamount]
&&\phantom{{}=-{1\over 16}\left[\sin^2{1\over 2}\left(
\theta_{\smash\bmalpha\vphantom\bmlambda}+
\theta_{\bmalpha+\bmlambda}\right)\right]^{-1/16}\bigg[}
+2\epsilon_2\cos{1\over 2}\left(
\theta_{\smash\bmalpha\vphantom\bmlambda}-
\theta_{\bmalpha+\bmlambda}\right)\cos{1\over 2}\left(
\theta_{\smash\bmalpha\vphantom\bmlambda}+
\theta_{\bmalpha+\bmlambda}\right)\bigg]\,,\nonumber\\
\label{a.10}
\end{eqnarray}
the trigonometric identities
\begin{equation}
\left.\begin{array}{l}
\cos x-\cos y=-2\sin{1\over 2}(x-y)\sin{1\over 2}(x+y)\\[\medskipamount]
\sin x-\sin y=2\sin{1\over 2}(x-y)\cos{1\over 2}(x+y)
\end{array}\right\}\label{a.11}
\end{equation}
being used in passing to the final form.
The complementary identities
\begin{equation}
\left.\begin{array}{c}
\cos x+\cos y=2\cos{1\over 2}(x-y)\cos{1\over 2}(x+y)\\[\medskipamount]
\sin x+\sin y=2\cos{1\over 2}(x-y)\sin{1\over 2}(x+y)
\end{array}\right\}\label{a.12}
\end{equation}
permit expression (\ref{a.10}) to be further simplified, and one sees
that, under (\ref{6.11}), the measure (\ref{a.7}) gets multiplied by
\begin{equation}
\sum_\bmalpha\left\{-\epsilon_1\sin\theta_\bmalpha+\epsilon_2
\cos\theta_\bmalpha-{1\over 16}\sum_\bmlambda\left[-\epsilon_1\left(
\sin\theta_{\smash\bmalpha\vphantom\bmlambda}+
\sin\theta_{\bmalpha+\bmlambda}\right)+\epsilon_2\left(
\cos\theta_{\smash\bmalpha\vphantom\bmlambda}+
\cos\theta_{\bmalpha+\bmlambda}\right)\right]\right\}\,.\label{a.13}
\end{equation}
But this vanishes by virtue of the fact that $\bmlambda$ ranges over
8 values per site.
It follows that the measure is invariant under (\ref{6.11}).
\subsection*{Proof of the existence of the functional integrals
in (\ref{6.8})}
The integrands become singular when two or more $\theta_\bmalpha$'s
approach simultaneously a common value~$\bar\theta$.
Consider therefore a rectangular subvolume of the lattice containing
$JKLM$ sites, where $J$, $K$, $L$, $M$ are integers between 1 and
$N$, at least one of which is greater than~1.
The number of links in this subvolume is $4JKLM-JKL-JKM-JLM-KLM$.
Replace the integration variables $\theta_\bmalpha$ attached to the
$JKLM$ sites by $\bar\theta$, $\Delta\theta$ and $JKLM-2$ angular
variables in \mbox{$(JKLM-1)$-dimensional} Euclidean space.
The variable $\Delta\theta$ is the maximum distance of any of the
$\theta_\bmalpha$'s from the common value $\bar\theta$ and is the
critical variable for determining the convergence of the associated
subintegration.
At small values of $\Delta\theta$ the $\Delta\theta$ subintegration goes
like
\begin{eqnarray}
\int_0(\Delta\theta)^{-{1\over 4}(4JKLM-JKL-JKM-JLM-KLM)}
      (\Delta\theta)^{JKLM-2}\,d(\Delta\theta)&&\nonumber\\[\medskipamount]
=\int_0(\Delta\theta)^{{1\over 4}(JKL+JKM+JLM+KLM)-2}
\,d(\Delta\theta)\,.&&\label{a.14}
\end{eqnarray}
But $JKL+JKM+JLM+KLM\ge 7$.
Therefore this subintegration is no more singular than
$\int_0(\Delta\theta)^{-1/4}\,d(\Delta\theta)$, which converges.
Since all the subintegrations converge, the integral for the whole
lattice converges.

The convergence is no accident.
All the $\theta_\bmalpha$'s on the whole lattice can be brought
arbitrarily close to one another by imposing a sufficiently powerful
boost (the headlight effect).
But the integration measure has been shown above to be boost invariant.

It is a curious fact that the measure (\ref{a.7}) is the same as the
measure that appears in the computation of the string amplitude for
$N^4$ open-string tachyons on the boundary of a disk.
It is a consequence of the $O(1,2)$ symmetry of both systems.
The above proof of convergence is the same as that used in string
theory.
\clearpage
\subsection*{Figure Captions}
\parindent0pt
{\bf Figure 1}.
Histograms of the cross-link differences $\Delta\theta=
\theta_{\smash\bmalpha\vphantom\bmlambda}-\theta_{\bmalpha+\bmlambda}$
for the exact lattice action (\ref{2.2}), at $\beta=10^{-4}$ and $N=10$,
and for the \mbox{$\theta$-part} of the approximate action
$S_{\omega,\theta}$ (eq.~(\ref{5.8})) at $N=10$.
All cross-link differences for 60,000 sweeps %-EAM%(2.4~billion data bits)
have been compiled into 101~bins.
The normalized distributions for the two cases are indistinguishable.

\vspace{\medskipamount}
{\bf Figure 2}.
Histogram of the $s_\bmalpha$ at $\beta=10^{-3}$ and $N=10$,
superimposed upon the theoretical Gaussian distribution derived from
eq.~(\ref{5.6}) and the \mbox{$\omega$-part} of the approximate
action~$S_{\omega,\theta}$.

\vspace{\medskipamount}
{\bf Figure 3}.
Histogram of the $s_\bmalpha$ at $\beta=10^{-4}$ and $N=10$,
superimposed upon the theoretical Gaussian distribution derived from
eq.~(\ref{5.6}) and the \mbox{$\omega$-part} of the approximate
action~$S_{\omega,\theta}$.

\vspace{\medskipamount}
{\bf Figure 4}.
Plots of $2/\bar K^2
\left<\tilde\sigma^i_{R\bf k}\tilde\sigma^i_{R\bf k}{}^*\right>$
vs.~$\bar K^2\!\big/\beta_R$ for various values of $\beta$, at $N=10$.
$\beta_R$~is determined for each curve from its intercept at
$\bar K^2=0$ as obtained from a linear fit to the low-momentum values.
The curves are plotted as functions of $\bar K^2\!\big/\beta_R$ because
since $\bar K^2\!\big/\beta_R$ becomes $p^2\!\big/\mu_R{}^2$ in the
continuum limit, this is the only scale-invariant way to compare them.
Although the curves are not perfectly smooth they show remarkably little
dependence on the detailed values of the components of~$\bf k$.
Their essential dependence on $\bar K^2$ alone shows that $O(4)$
invariance is maintained in the continuum limit.
(See the discussion following eq.~(\ref{1.15}).)
The error bars (statistical plus autocorrelation) range from .1\% at
the high-momentum end to 1\% at the low-momentum end.
Only the latter are shown on the figure.

For $\beta\le 10^{-3}$ the curves are seen to fall on top of one another
at low momenta.
The difference between the curves at higher momenta is explained by the
spikiness of $\left<\tilde\sigma^i_{Rk}\tilde\sigma^i_{Rk}{}^*\right>$
arising from $\Omega(\bmalpha-\bmalpha')$ in eq.~(\ref{7.1}).
The spikiness is a short-wavelength phenomenon that is picked up by the
Fourier transform.
It goes away as $\beta\to 0$.
The $\beta=0$ curve itself is obtained by using the \mbox{$\theta$-part}
of the action~$S_{\omega,\theta}$.

\vspace{\medskipamount}
{\bf Figure 5}.
A plot of $\beta_R$ vs.~$\beta$ as determined from the intercepts of the
curves shown in Fig.~4.
The error bars, which are less than a tenth of a percent, are almost too
small to be seen.
At higher values of $\beta$ the model tends toward a free-field model,
and $\beta_R\to\beta$.
It is clear that there is no value of $\beta$ for which $\beta_R$
vanishes.

\vspace{\medskipamount}
{\bf Figure 6}.
A plot of $\beta_R$ vs.~$N$ when $\beta=0$, obtained using the
\mbox{$\theta$-part} of the action~$S_{\omega,\theta}$.
The asymptotic value of $\beta_R$, as $N\to\infty$, is approximately
.078\,.

\vspace{\medskipamount}
{\bf Figure 7}.
Plots of $\beta_R$ vs.~$\beta$ at $N=10$.
The black circles represent the values obtained from the $\sigma$
\mbox{2-point} function.
The triangles represent the values obtained from current-current
averages at low momenta.
Error bars are shown only for the triangles.
Error bars for the circles are too small to be seen.
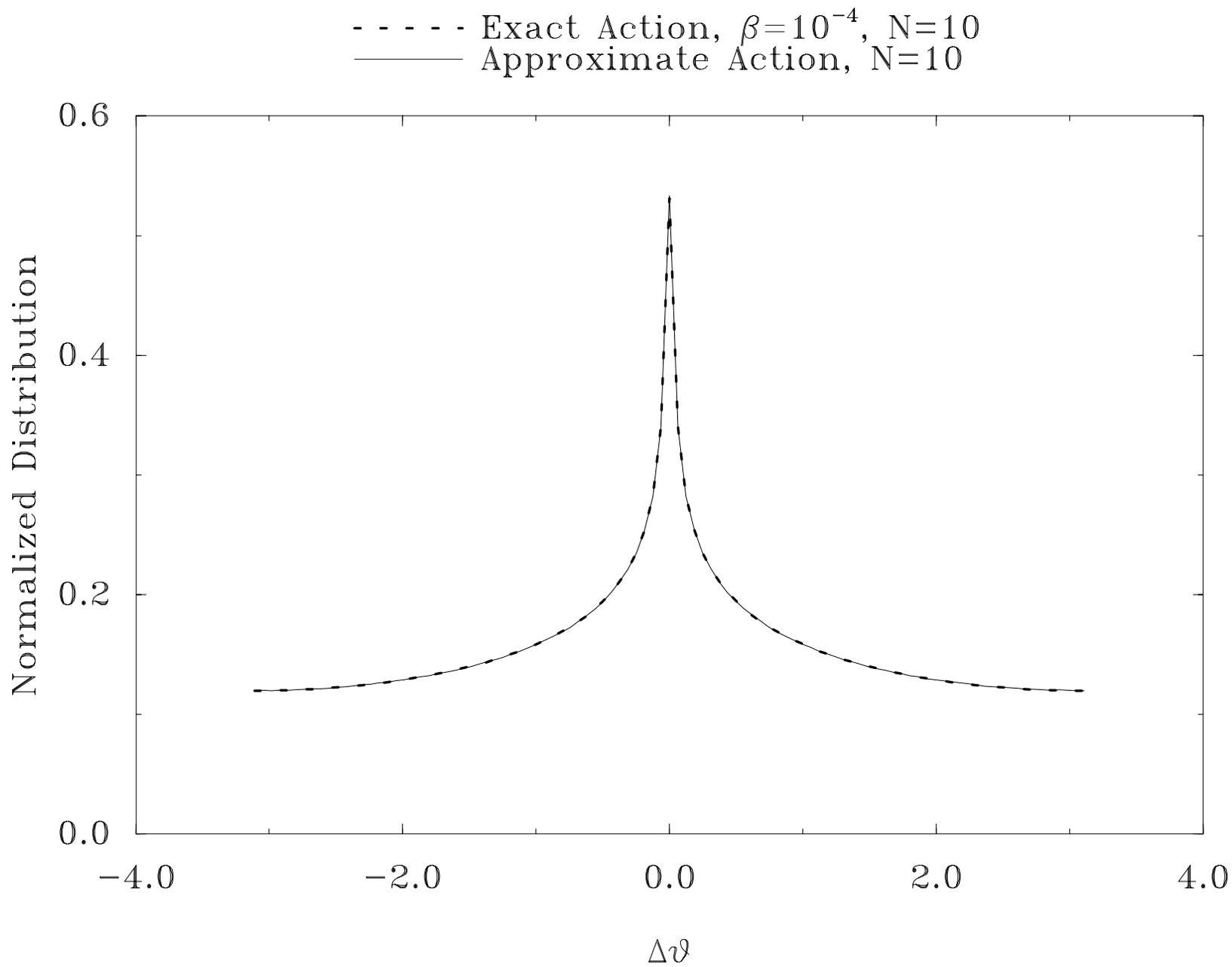
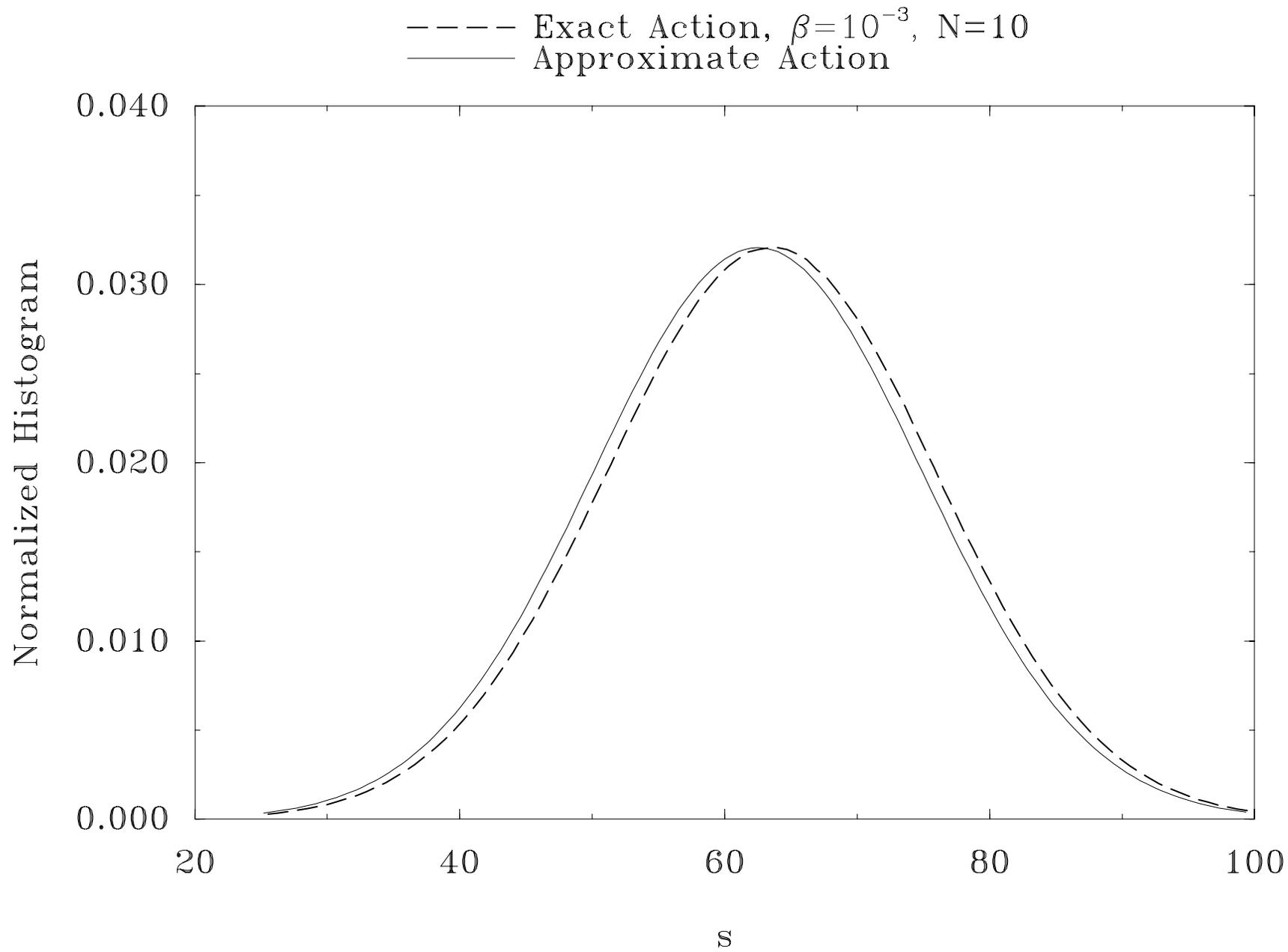
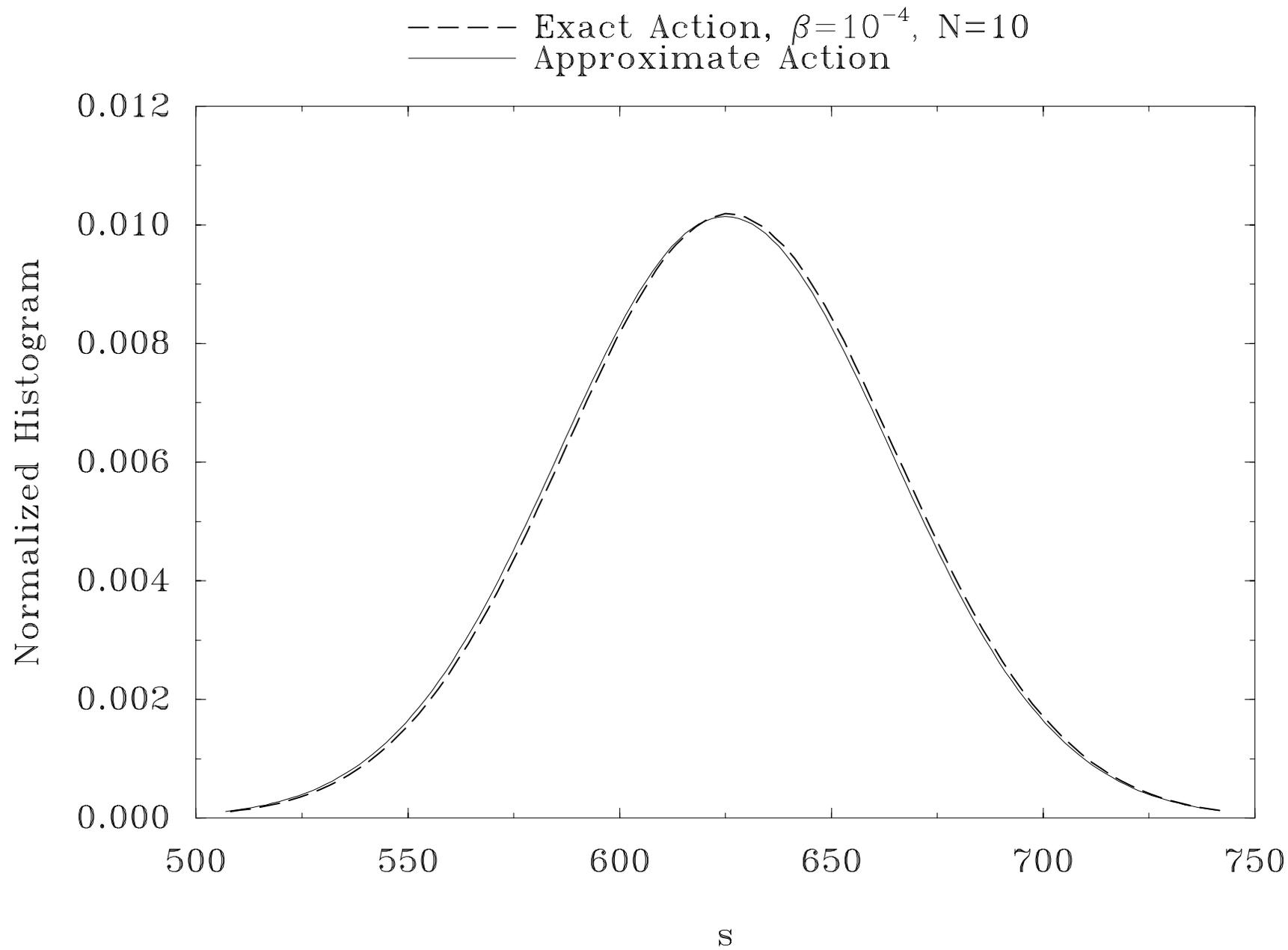
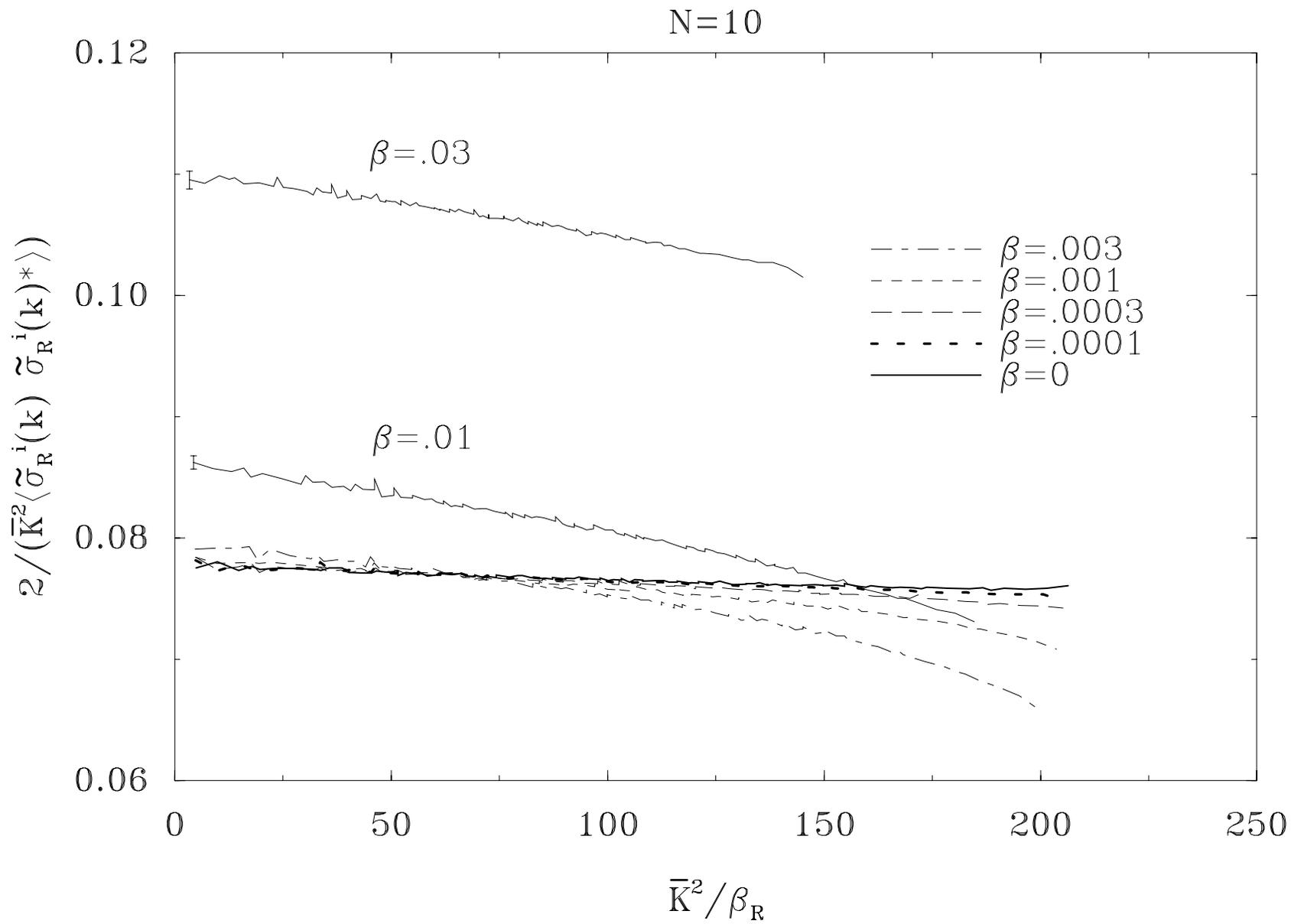
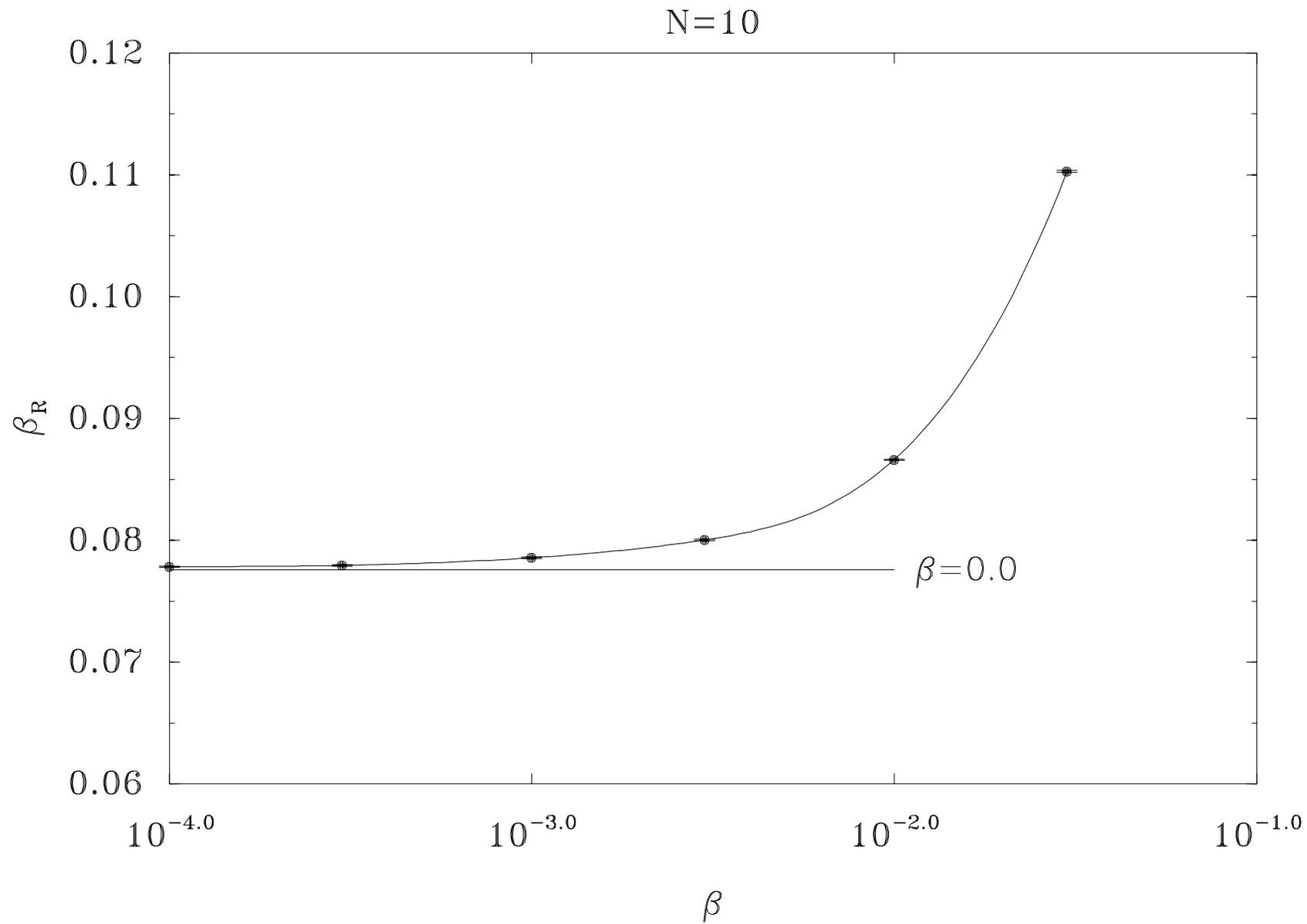
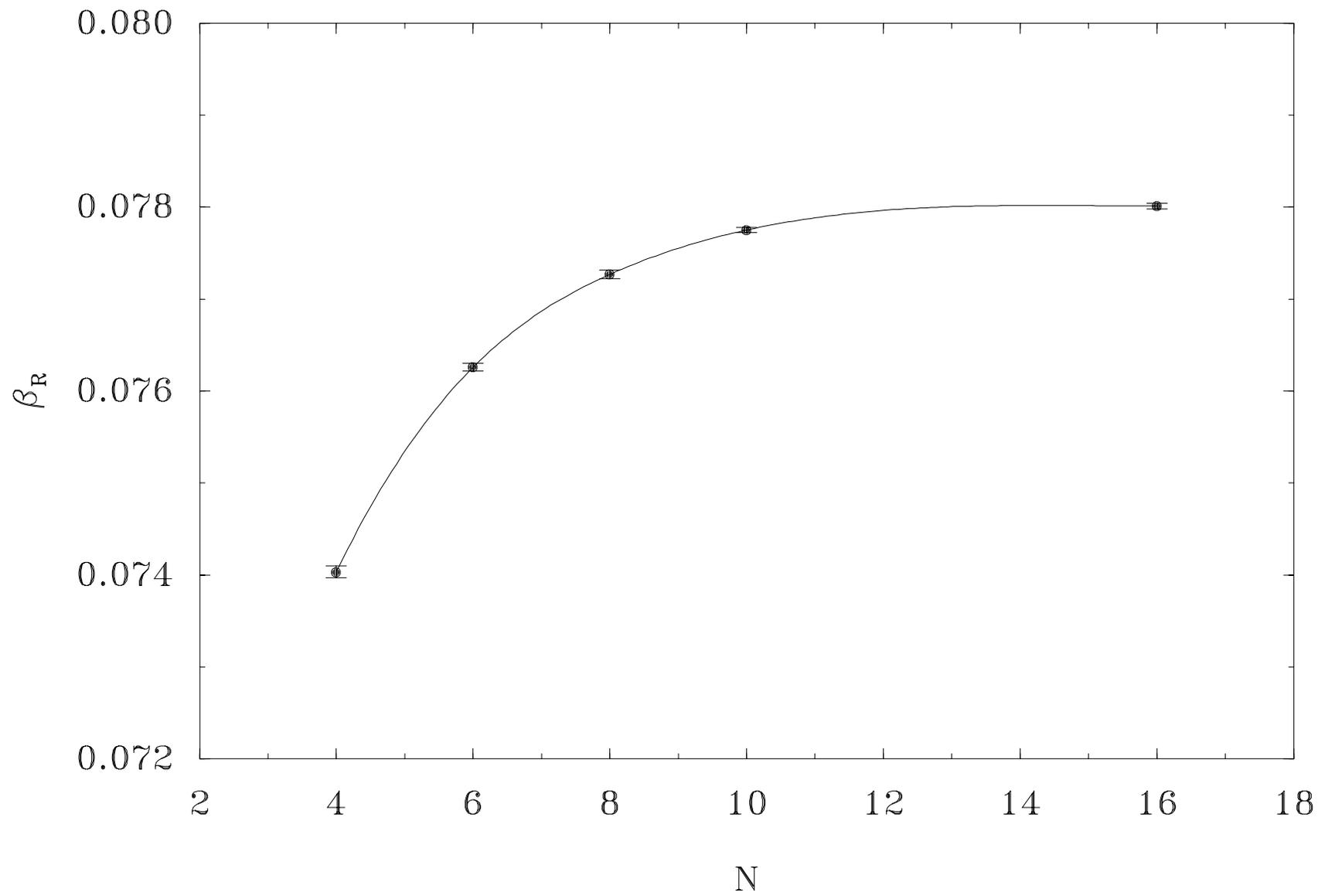
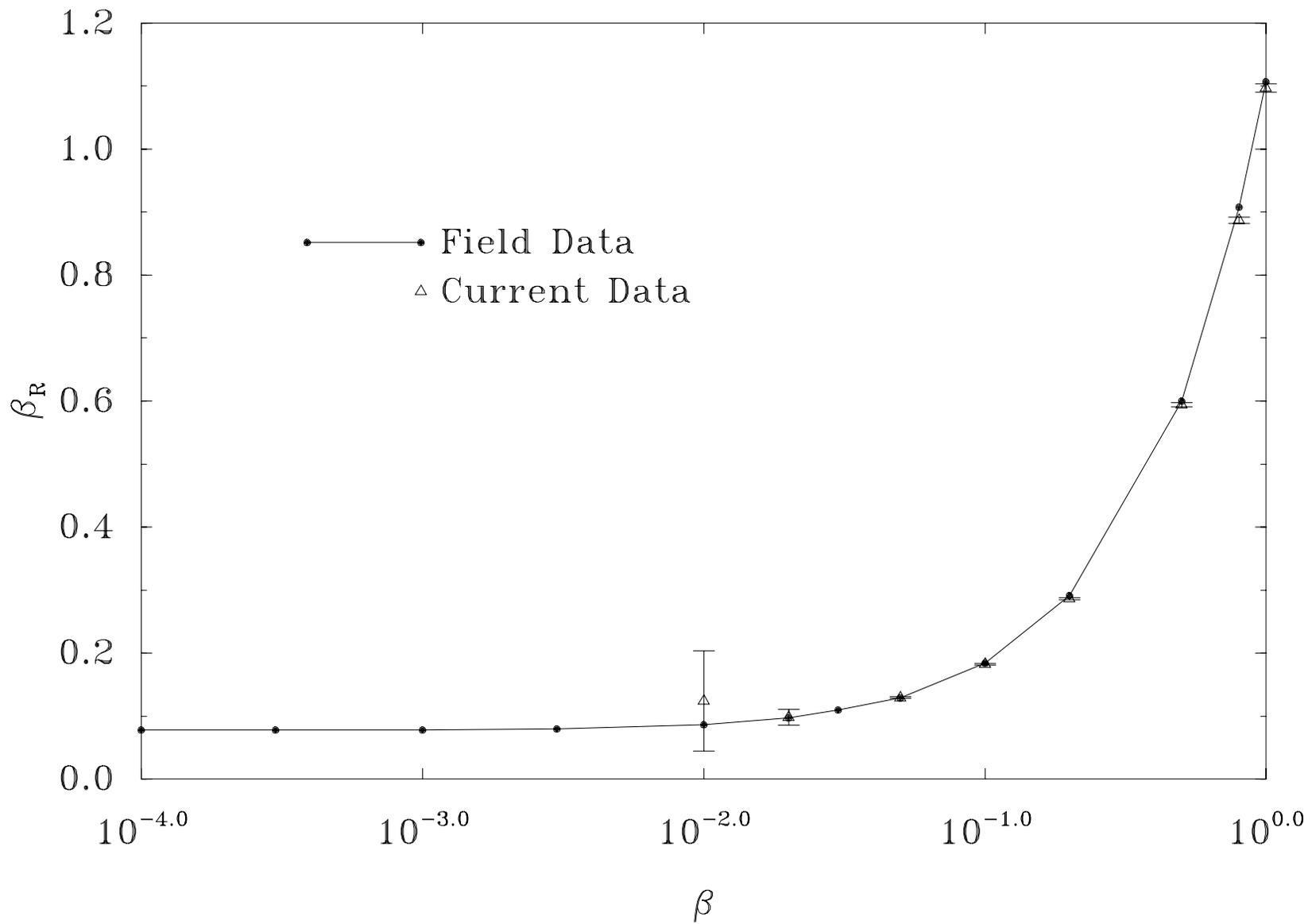
\end{document}